\documentclass[%
 reprint,
nofootinbib,
 amsmath,amssymb,
 aps,
]{revtex4-2}

\usepackage{graphicx}
\usepackage{dcolumn}
\usepackage{bm}
\usepackage{bbm}
\usepackage[caption=false]{subfig}
\usepackage[colorlinks=true, allcolors=blue]{hyperref}

\newcommand{\dd}{\mathrm{d}}
\newcommand{\ii}{\mathrm{i}}

\newcommand{\GF}{G_{\mathrm{F}}}
\newcommand{\abs}[1]{\left\lvert#1\right\rvert}
\newcommand{\norm}[1]{\left\lvert\left\lvert#1\right\rvert\right\rvert}
\newcommand\barparena[1]{\overset{\scriptscriptstyle(-)}{#1}}

\begin{document}

\title{Collisional dynamics of neutrino flavor conversion: angular structures, inhomogeneity, and decoherence}

\author{Masamichi Zaizen}
\affiliation{Department of Earth Science and Astronomy, The University of Tokyo, Tokyo 153-8902, Japan}
\email{mzizn@g.ecc.u-tokyo.ac.jp}

\date{\today}

\begin{abstract}
Self-interacting neutrino gases exhibit collective behavior, in which additional incoherent collisions can induce collisional neutrino-flavor instability.
We investigate the effects of energy dependence, anisotropy, and spatial inhomogeneity on the resulting flavor conversion through linear stability analysis and nonlinear quantum kinetic simulations.
We find contrasting wavenumber dependences for the near/gapless and far/gapped branches, with maximum growth rates at collision-scale shifted wavenumbers and at collective-scale wavenumbers, respectively.
Their eigenvectors reveal distinct energy and angular structures, which are reflected in the subsequent establishment of flavor swap and equipartition.
The nonlinear evolution also features secondary flavor growth associated with changes in the stability of collective modes as the neutrino background evolves.
Despite the spreading of flavor coherence across spatial scales due to nonlinear mode coupling, collisional decoherence tends to limit the development of spatial inhomogeneities, leaving the neutrino populations nearly homogeneous.
These findings represent a step towards understanding and predicting the nonlinear flavor evolution of dense neutrino gases interacting with their surroundings, particularly in explosive astrophysical environments.
\end{abstract}

\maketitle


\section{Introduction}
In dense neutrino gases, neutrino self-interactions give rise to nonlinear flavor dynamics, called collective neutrino oscillations \cite{Pantaleone:1992,Sigl:1993,Duan:2010,Richers:2022b,Tamborra:2021,Johns:2025,Fiorillo:2026h}.
The collective motions have been studied using neutrino dispersion relations by analogy with the plasma physics, revealing the existence of various flavor instabilities \cite{Sawyer:2005,Sawyer:2016,Izaguirre:2017,Johns:2023a}.
Different instability mechanisms can lead to distinct final flavor states.

Such extreme environments are realized in explosive astrophysical phenomena, such as core-collapse supernovae (CCSNe) and binary neutron-star merger (BNSM) remnants \cite{Fischer:2024,Volpe:2024,Tamborra:2025,Raffelt:2026}.
The occurrence of flavor conversion can drastically change the neutrino radiation field and could quantitatively alter the fluid dynamics and observables \cite{Just:2022,Li:2021,Fernandez:2022,Lund:2025,Qiu:2025,Qiu:2025a,Ehring:2023,Ehring:2023a,Ehring:2026,Nagakura:2024,Mori:2025,Wang:2025,Wang:2025a,Wang:2025b,Akaho:2026,Gogilashvili:2026,Gogilashvili:2026a}.
Thereby, the rich nonlinear behaviors have attracted considerable attention resolving uncertainties within the quantum regime in neutrino transport.

Self-induced flavor conversion, called fast flavor conversion (FFC), has been widely investigated due to its {\it fast} growth on the self-interaction timescale \cite{Sawyer:2005,Sawyer:2016,Izaguirre:2017}.
FFC is triggered just by the imbalance in the neutrino angular distributions on flavor isospin space and evolves on timescale by the self-interactions.
In the nonlinear dynamics, the spatial mode coupling through the advection term brings the development of a cascade, leading to a flavor depolarization \cite{Bhattacharyya:2021,Richers:2021,Zaizen:2021a,Liu:2026a}.
Such dynamics have been observed as the asymptotic behaviors towards flavor equipartition in local simulations and motivated the development of the analytical models without direct quantum kinetic simulations \cite{Wu:2021,Bhattacharyya:2022,Zaizen:2023,Xiong:2023a,Abbar:2024a,Richers:2024,Goimil-Garcia:2025a}.

On the other hand, collision-induced flavor conversion, called collisional flavor conversion (CFC), has attracted attention because of its distinct driving mechanism \cite{Johns:2023a} and the occurrence region in astrophysical environments \cite{Liu:2023a,Akaho:2024a,Nagakura:2025a,Froustey:2026,Urquilla:2026a,Ng:2026}.
CFC is induced by the disparity in the flavor-dependent {\it collisional decoherence}.
In such dense surrounding environments, neutrino angular distributions are relatively isotropized due to the non-negligible collisions.
Thereby, most linear and nonlinear studies have focused on isotropic and homogeneous neutrino gases and have nevertheless revealed rich dynamics \cite{Padilla-Gay:2022a,Lin:2023,Xiong:2023b,Liu:2023,Kato:2024,Johns:2023c,Zaizen:2025,Froustey:2025}.
One such phenomenon is resonancelike enhancement as the net lepton number density approaches zero $N_{\nu} \sim N_{\bar{\nu}}$.
In the resonancelike regime, the growing scale is coupled to the self-interaction strength besides collisions, resulting in a flavor swap \cite{Kato:2024}.
The nontrivial behaviors are sustained even in the anisotropic neutrino background intrinsically leading to FFC and exhibit more complicated flavor evolution episodes \cite{Takahashi:2026}.

In the non-resonancelike regime, unstable branches in the dispersion relation are classified into the plus/near/gapless and minus/far/gapped modes, with distinct eigenfrequencies \cite{Lin:2023,Liu:2026,Fiorillo:2026c}.
Based on the isotropic, monochromatic, and spatially homogeneous assumptions, the approximate solutions show that they have collision-scaled (or negligible) and collective-scaled real parts of the eigenfrequencies, respectively.
This distinction is reflected in the corresponding eigenvector structures in the linear regime, leading to the distinct asymptotic states: flavor swap and equipartition \cite{Johns:2023c,Zaizen:2025,Froustey:2025}.

However, the effects of neutrino anisotropy and spatial inhomogeneity have received relatively little attention.
Surveys of flavor instabilities in explosive astrophysical environments have implied that CFC may occur in the semi-transparent domain with matter density $10^{10-12}\mathrm{\,g\,cm^{-3}}$ \cite{Liu:2023a,Akaho:2024a,Nagakura:2025a,Froustey:2026,Urquilla:2026a,Ng:2026}, particularly with density reaching approximately $10^{13}\mathrm{\,g\,cm^{-3}}$ near the neutrino energy sphere for the resonancelike type \cite{Akaho:2024a,Liu:2024,Ng:2026}.
The high matter density makes neutrino distributions more isotropic due to scattering and charged-current reactions overwhelming outward advection.
The presence of flavor-dependent neutrino spheres could nevertheless allow the emergence of anisotropic neutrino gases in the non-resonancelike regions, particularly in the outer layer.
This motivates us to investigate the impact or response from the anisotropy and spatial inhomogeneity to the collisional dispersion relation.
In particular, how does a forward-peaked neutrino angular distribution affect the wavenumber dependence of collisional neutrino instabilities away from the zero mode?

In this work, we analyze collisional neutrino flavor instability in a multi-energy, multi-angle, and spatially-inhomogeneous neutrino background through a dispersion relation approach and then show the subsequent nonlinear flavor evolution.
Our study highlights the role of anisotropic structures and finite wavenumber modes and clarifies the competition between the development of a spatial cascade and collisional decoherence, which have both been neglected in many previous papers.

This paper is organized as follows.
Sec.\,\ref{Sec:2_QKE} describes quantum kinetic equations and our setups leading to two kinds of unstable branches dominated in multi-energy and anisotropic neutrino gases.
In Sec.\,\ref{Sec:3_Linear}, we present how the collisional modes behave according to the dispersion relations in the multi-energy, anisotropic, and spatially-inhomogeneous neutrino background.
We then examine the nonlinear flavor evolution, including the exponential growth, collisional decoherence, and flavor depolarization in Sec.\,\ref{Sec:4_Nonlinear}.
Finally, we conclude and summarize our results in Sec.\,\ref{Sec:5_summary}.
Throughout the paper, we use natural units with $\hbar = c = 1$ and a metric with $(+,-,-,-)$.

\section{Quantum Kinetic Equation}\label{Sec:2_QKE}
Flavor evolution within self-interacting neutrino gases is governed by the quantum kinetic equation (QKE):
\begin{equation}
    \left(p^{\mu}\frac{\partial}{\partial x^{\mu}} - \Gamma^{j}_{\alpha\beta}p^{\alpha}p^{\beta}\frac{\partial}{\partial p^j}\right)\barparena{\rho} = -\ii\left[\barparena{\mathcal{H}}_{\mathrm{osc}},\,\barparena{\rho}\right] + \mathcal{C}\left[\barparena{\rho}\right],
\end{equation}
where the barred symbols denote the anti-neutrino sectors.
In local Cartesian coordinates in flat spacetime, the connection coefficients $\Gamma^j_{\alpha\beta}$ vanish in the transport equations.
The oscillation term is given by
\begin{align}
    \barparena{\mathcal{H}}_{\mathrm{osc}} = \mathcal{H}_{\nu\nu} = \sqrt{2}\GF\,v^{\alpha}\int\dd\Gamma^{\prime}\,v_{\alpha}^{\prime}\rho^{\prime},
\end{align}
where we drop the vacuum and matter terms because this paper focuses solely on the collision-induced flavor instability.
Also, the neutrino density matrix follows the flavor-isospin convention $\bar{\rho}(E) = -\rho(-E)$ with negative occupation for antineutrinos.
Collision terms with opacity $\kappa_{\alpha}$ for a flavor $\alpha$ within quantum kinetics are given in the following relaxation form
\begin{equation}
    \mathcal{C}[\rho] = \frac{1}{2}\left\{\begin{pmatrix}
        \kappa_e & 0 \\ 0 & \kappa_x
    \end{pmatrix},\, \rho^{\mathrm{eq}} - \rho\right\},
\end{equation}
where $\rho^{\mathrm{eq}}$ denotes the corresponding equilibrium state.
This term is decomposed into the classical diagonal parts and the quantum off-diagonal:
\begin{align}
\mathcal{C}[\rho] &= \begin{pmatrix}
    \kappa_e\left(\rho^{\mathrm{eq}}_{ee}-\rho_{ee}\right) & 0 \\ 0 & \kappa_x\left(\rho^{\mathrm{eq}}_{xx}-\rho_{xx}\right)
\end{pmatrix} \notag \\
    &\qquad+ \begin{pmatrix}
    0 & -R_{\mathrm{EA}}\rho_{ex} \\ -R_{\mathrm{EA}}\rho_{xe} & 0
\end{pmatrix} \\
    &= \mathcal{C}_{\mathrm{cls}} + \mathcal{C}_{\mathrm{qke}},
\end{align}
where $R_{\mathrm{EA}} = (\kappa_e+\kappa_x)/2$.
The first part $\mathcal{C}_{\mathrm{cls}}$ corresponds to a flavor population changing process in the classical transport and describes  relaxation towards the classical equilibrium, while the second $\mathcal{C}_{\mathrm{qke}}$ provides quantum decoherence that can lead to a collision-induced flavor instability.
For simplicity, we take $\kappa_x = 0$ ignoring collision processes involving heavy-leptonic flavors and neglect the diagonal collision term $\mathcal{C}_{\mathrm{cls}}$, thereby excluding relaxation of flavor populations towards quantum equilibrium.

In this study, we simulate flavor evolution on multi-energy, multi-angle, and spatially-inhomogeneous neutrino gases $\rho = \rho(t;z;E_{\nu},v_z)$.
The energy distribution of flavor $\alpha$ is given by
\begin{equation}
    f_{\nu_{\alpha}}(E_{\nu}) \propto \frac{1}{\exp[(E_{\nu}-\mu_{\nu_{\alpha}})/T_{\nu_{\alpha}}] + 1}.
\end{equation}
We adopt the parameters used in our previous study \cite{Zaizen:2025}; we set chemical potentials $\mu_{\nu_{\alpha}}$ to zero and instead impose the ratio of flavor populations $\bar{n}_{\nu_e}/n_{\nu_e} = 0.8$, and the neutrino temperatures are $(T_{\nu_e},T_{\bar{\nu}_e}) = (4\mathrm{\,MeV}, 5\mathrm{\,MeV})$.
To investigate the impact of anisotropy of neutrino background, we employ forward-peaked Gaussian distributions \cite{Zaizen:2023}:
\begin{equation}
    g_{\nu_{\alpha}}(v_z) \propto \exp\left[-(v_z-1)^2 / 2\xi_{\nu_\alpha}^2\right]
\end{equation}
with $\xi_{\nu_e} = 0.6$ and $\xi_{\bar{\nu}_e} = 0.53$.
There is then no angular crossing in that setup for neutrino number densities.
We here choose the angular distributions which are intentionally more strongly forward-peaked than those expected in realistic astrophysical environments to illustrate the effects of anisotropy more clearly.

We solve the QKE in a one-dimensional local box using the {\tt GANTS-QK} code \cite{Zaizen:2026}, where all quantities are normalized by the self-interaction strength defined by $\mu_{\nu_e} = \sqrt{2}\GF\,n_{\nu_e} \equiv 1$.
Then, time $t$ and space $z$ are expressed in dimensionless form with the units of $\mu^{-1}$.
We simulate the flavor evolution in a local box $L_z = 1000$ with the grid size $N_z = 1000$.
And the resolution for momentum space is set by $N_E = 100$ in $E_{\nu} \in [0,100]\mathrm{\,MeV}$ and $N_v = 256$, respectively.
To investigate the collision-induced flavor conversion, we adopt the energy-dependent reaction rates:
\begin{equation}
    \barparena{R}_{\mathrm{EA}}(E_{\nu}) = \barparena{R}_0(E_{\nu})\left(\frac{E_{\nu}}{10\mathrm{\,MeV}}\right)^2,
\end{equation}
where $R_0 = 10^{-4}$ in the neutrino sector, and we consider two reaction rate configurations $\bar{R}_0 = 10^{-5}$ and $10^{-4}$ in the antineutrino sector.
The former case satisfies $\langle R\rangle > \langle\bar{R}\rangle$ for energy-averaged reaction rates and exhibits that the near/gapless mode \cite{Liu:2026,Fiorillo:2026c} is unstable (previously called the plus mode in the previous studies based on the isotropic and homogeneous background \cite{Lin:2023,Liu:2023,Zaizen:2025,Froustey:2025}).
On the other hand, the latter case satisfies $\langle R\rangle < \langle\bar{R}\rangle$, where the far/gapped mode is dominant (previously called the minus mode)\footnote{The originally called ``plus/minus'' was named with the sign in the analytical solutions \cite{Lin:2023,Liu:2023}, whereas the ``near/far'' or ``gapless/gapped'' stems from the real part of wave frequency $\omega$.
The enhancement (or cancelling) due to the sign determines the scale of the eigenfrequency.
When the unstable branch is within the collective scale $\mathrm{Re}\,\omega \sim \mu$, it is classified into the far/gapped mode, and vice versa.}.

\section{Collisional Modes}\label{Sec:3_Linear}
In order to capture the nonlinear dynamics, it is important, as a first step, to explore the unstable branch responsible for flavor instabilities and the corresponding eigenvectors within the linear regime.

\subsection{Methodology}
In the linear limit, we assume sufficiently small flavor coherence $\abs{\rho_{ex}}\ll\abs{\rho_{ee}}$ and neglect the wave-wave (nonlinear) coupling.
We adopt a plane-wave ansatz with $K^{\mu}\equiv(\Omega,\boldsymbol{K})$ for the off-diagonal components
\begin{equation}
    \rho_{ex}(t,z; E,v) \propto \tilde{Q}_{E,v,\Omega,K}\exp\left[-\ii K^{\mu}x_{\mu}\right].
\end{equation}
The governing equation for flavor coherence can be then rewritten as
\begin{align}
    \left(\Omega - \boldsymbol{v}\cdot\boldsymbol{K}\right)\tilde{Q} &= \sqrt{2}\GF\, \tilde{Q}\,v^{\mu}\int\dd\Gamma^{\prime}v_{\mu}^{\prime}(\rho_{ee}^{\prime}-\rho_{xx}^{\prime}) \notag \\
        & - \sqrt{2}\GF\,(\rho_{ee}-\rho_{xx})v^{\mu}\int\dd\Gamma^{\prime}\,v_{\mu}^{\prime}\tilde{Q}^{\prime} \notag\\
        & - \ii R_{\mathrm{EA}}\tilde{Q}.
\end{align}
Introducing shifted wave frequency $k^{\mu} \equiv K^{\mu} - \Phi^{\mu}$ using the lepton-current potential and a flavor-difference distribution $G_{ex}(E_{\nu},v) \equiv \sqrt{2}\GF(\rho_{ee}-\rho_{xx})$, the equation becomes 
\begin{align}
    (v^{\nu}k_{\nu} + \ii R_{\mathrm{EA}})\tilde{Q} &= - G_{ex} v^{\mu}\int\dd\Gamma^{\prime}\, v_{\mu}^{\prime} \tilde{Q}^{\prime} \\
        &\equiv G_{ex} v^{\mu}\tilde{a}_{\mu}.
\end{align}
The corresponding eigenfunction is then given by
\begin{equation}
    \tilde{Q}_{E,v,\omega,k} = G_{ex}(E_{\nu},v) \frac{v^{\mu}\tilde{a}_{\mu}}{v^{\nu}k_{\nu} + \ii R_\mathrm{EA}},
    \label{eq:eigenvec}
\end{equation}
and eliminating $\tilde{Q}$, we obtain the following equation
\begin{equation}
    v\cdot\tilde{a} = -\, v^{\mu}\int\dd\Gamma^{\prime}\,G_{ex} ^{\prime}v_{\mu}^{\prime}\frac{v\cdot\tilde{a}}{v\cdot k + \ii R_{\mathrm{EA}}}
\end{equation}
Defining the polarization tensor,
\begin{equation}
    \Pi^{\mu\nu}(\omega,\boldsymbol{k}) \equiv \eta^{\mu\nu} + \int\dd\Gamma\,G_{ex}\frac{v^{\mu}v^{\nu}}{v\cdot k + \ii R_{\mathrm{EA}}},
\end{equation}
the linearized equation is summarized as
\begin{equation}
    \Pi^{\mu\nu}\tilde{a}_{\nu} = 0.
    \label{eq:DR_mat_eq}
\end{equation}
The nontrivial solution requires
\begin{equation}
    \mathcal{D}(\omega,k) \equiv \mathrm{det}\left[\Pi^{\mu\nu}(\omega,\boldsymbol{k})\right] = 0.
    \label{eq:determinant_D}
\end{equation}
In a one-dimensional system with axial symmetry along the $z$-axis, the polarization tensor is 
\begin{equation}
    \Pi^{\mu\nu}(\omega,\boldsymbol{k}) = \begin{pmatrix}
        I_0 + 1 & 0 & 0 & I_1 \\
        0 & \frac{I_0 - I_2}{2} - 1 & 0 & 0 \\
        0 & 0 & \frac{I_0 - I_2}{2} - 1 & 0\\
        I_1 & 0 & 0 & I_2 - 1
    \end{pmatrix},
\end{equation}
where
\begin{equation}
    I_n \equiv \int\dd\Gamma\,G_{ex}\frac{v^n}{v\cdot k + \ii R_{\mathrm{EA}}}.
    \label{eq:I_n_DR}
\end{equation}
The determinant condition in Eq.\,\eqref{eq:determinant_D} is translated into
\begin{equation}
    \left[\frac{I_0-I_2}{2}-1\right]^2\left[(I_0+1)(I_2-1) - I_1^2\right] = 0,
    \label{eq:DR_det}
\end{equation}
where a positive (negative) imaginary part of the wave eigenfrequency corresponds to exponential growth (damping) of perturbation in flavor coherence in time.

\subsection{Classification of collective modes}
For the axisymmetric neutrino background, Eq.\,\eqref{eq:DR_det} separates into {\it axial-symmetry breaking} and {\it axial-symmetry preserving branches} \cite{Raffelt:2013}.
The transverse sector satisfies
\begin{equation}
    I_0 - I_2 - 2 = 0
\end{equation}
and admits perturbations depending on the azimuthal angle.
In this work, we restrict our analysis to azimuthally symmetric perturbations and thereby consider only the longitudinal and axial-symmetry preserving branch,
\begin{equation}
    (I_0+1)(I_2-1) - I_1^2 = 0.
    \label{eq:det_branch}
\end{equation}

For the isotropic and spatially homogeneous background, the angular integrals satisfy $I_1 = 0$ and $I_2 = I_0/3$, and Eq.\,\eqref{eq:det_branch} reduces to
\begin{equation}
    (I_0 + 1)(I_0 - 3) = 0.
\end{equation}
The conditions $I_0 = 3$ and $I_0 = -1$ correspond to {\it isotropy-breaking}
\footnote{The former remains axisymmetric within the longitudinal sector: breaking the isotropy does not necessarily imply breaking the axial symmetry.}
and {\it isotropy-preserving branches} \cite{Liu:2023}.
In the presence of the anisotropy, $I_1$ is generally nonzero, coupling the longitudinal spatial and temporal components.
The resulting eigenmodes therefore generally mix angular structures, which are decoupled in the isotropic and homogeneous limit.

In the isotropic and homogeneous limit, $G_{ex}=G_{ex}(E_{\nu})$, each component of $\tilde{a}^{\mu}$ is given from Eq.\,\eqref{eq:DR_mat_eq} by
\begin{align}
\begin{cases}
    (I_0+1)\tilde{a}_0 = 0 \\
    (I_2-1)\tilde{a}_z = 0.
\end{cases}
\end{align}
The isotropy-preserving branch $I_0 = -1$ requires a nonzero $\tilde{a}_0$ and $\tilde{a}_z = 0$.
The eigenvector then has a structure depending only on energy
\begin{equation}
    \tilde{Q}(E_{\nu}) = G_{ex}(E)\frac{\tilde{a}_0}{\omega + \ii R_{\mathrm{EA}}},
\end{equation}
and the eigenvector retains the isotropic angular structure.
On the other hand, the breaking branch with $I_2 = I_0/3 = 1$ has $\tilde{a}_0 = 0$ and a nonzero $\tilde{a}_z$.
The resultant eigenvector has an angular dependent structure
\begin{equation}
    \tilde{Q}(E_{\nu},v) = G_{ex}(E)\frac{-v\tilde{a}_z}{\omega + \ii R_{\mathrm{EA}}},
\end{equation}
which violates the isotropy of the perturbation.

Considering spatial inhomogeneity $k\neq 0$, the isotropy-preserving branch acquires angular structure stemming from the denominator even for the isotropic neutrino background:
\begin{equation}
    \tilde{Q}(E_{\nu},v) = G_{ex}(E)\frac{\tilde{a}_0}{\omega - kv + \ii R_{\mathrm{EA}}}.
\end{equation}
Still, for intrinsically isotropic neutrino gases, the linearized system preserves parity symmetry $(v,k) \leftrightarrow (-v,-k)$ even in the presence of spatially inhomogeneous perturbations.
Consequently, the dispersion relation looks symmetric in $k$.

In the presence of anisotropy, $G_{ex}=G_{ex}(E_{\nu},v)$, a non-vanishing $I_1$ couples the symmetry-breaking branch to the preserving one as mentioned above and consequently yields the mixed eigenmodes,
\begin{equation}
    \tilde{Q}(E_{\nu},v) = G_{ex}(E,v)\frac{\tilde{a}_0 - \tilde{a}_z v}{\omega - kv + \ii R_{\mathrm{EA}}}.
\end{equation}
The eigenvector could have a root at $v = \tilde{r} \equiv \tilde{a}_0/\tilde{a}_z$.
This condition occurs only when $\tilde{r}$ is purely real and lies within the physical angular range.
Considering the amplitude of the eigenvector, the numerator is exactly recast into 
\begin{equation}
    \abs{\tilde{a}_0 - \tilde{a}_z v}^2 = \abs{\tilde{a}_z}^2\left[ (\mathrm{Re}\,\tilde{r}-v)^2 + (\mathrm{Im}\,\tilde{r})^2 \right].
    \label{eq:root_eigenvec}
\end{equation}
A nonzero imaginary part of the ratio $\tilde{r}$ behaves as the off-set in the angular amplitude.
And when the real part is outside the angular range $v\in [-1,1]$, the numerator monotonically varies in the angular directions.
These angular structures are illustrated as the demonstrations shown in Fig.\,\ref{fig:DR_eigen}.

Note that the distinction between isotropy-breaking and preserving branches is independent of the classification into near/gapless and far/gapped modes \cite{Liu:2026,Fiorillo:2026c}.
Each angular sector can possess both the collective-scale (self-interaction scale) roots and collision-scale roots, depending on the neutrino number densities and collision rates.
Consequently, the two components can support more than two frequency roots.
In the monochromatic and homogeneous limit, Eq.\,\eqref{eq:det_branch} is a quartic equation for $\omega$ because $I_n$ has both neutrino and anti-neutrino sectors with $\omega^{-1}$.
Actually, Eq.\,\eqref{eq:I_n_DR} in the limit reduces to
\begin{equation}
    I_n = \frac{A_n}{\omega + \ii R_{\mathrm{EA}}} - \frac{B_n}{\omega + \ii \bar{R}_{\mathrm{EA}}},
\end{equation}
where $A_n$ and $B_n$ denote $n-$th angular moments for neutrinos and anti-neutrinos.
The denominators yield a quadratic polynomial in $\omega$, and Eq.\,\eqref{eq:det_branch} yields a fourth-degree polynomial in $\omega$ in the generic case.

\begin{figure*}[t]
    \centering
    \includegraphics[width=1.\linewidth]{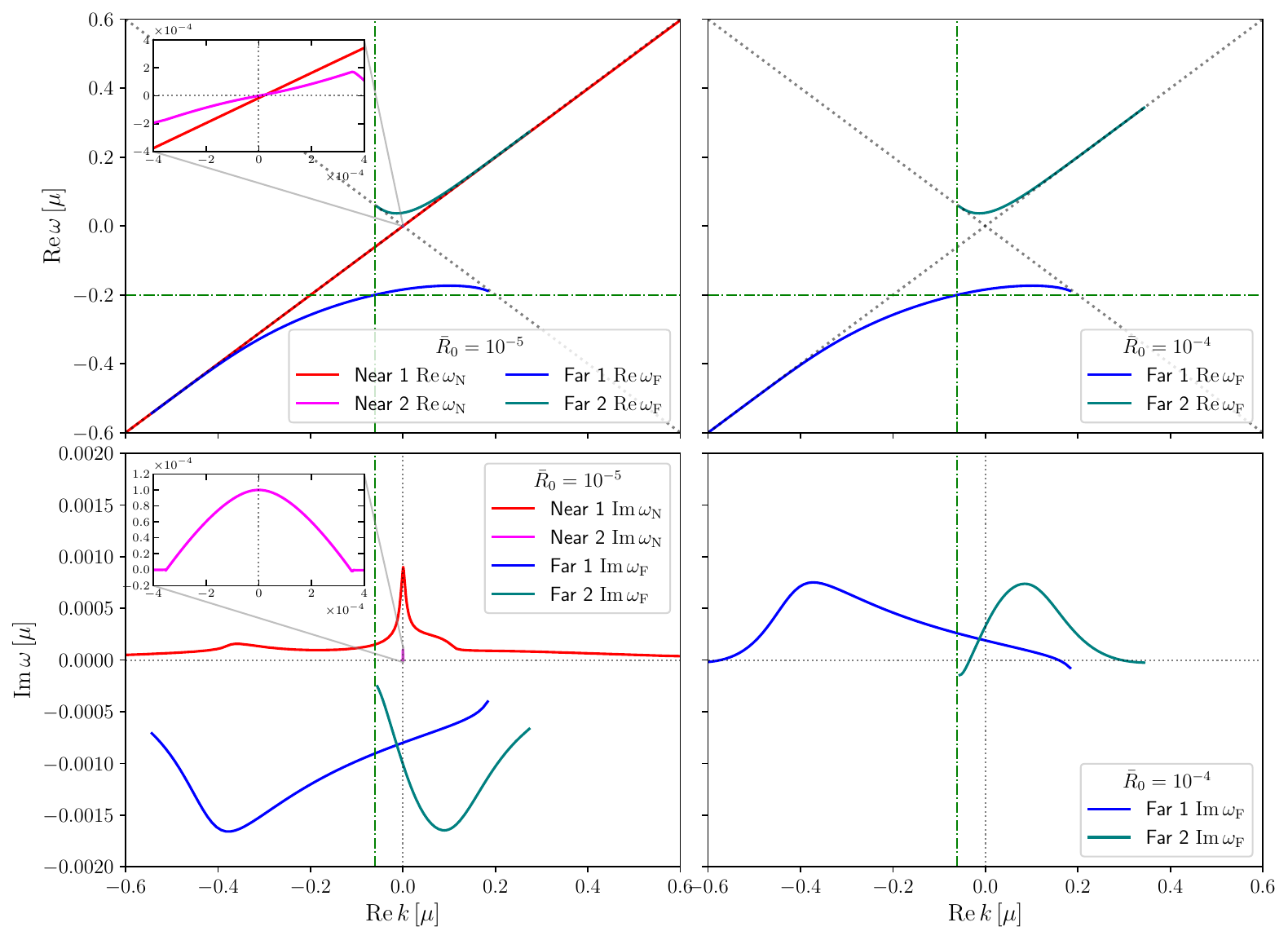}
    \caption{Real (top) and imaginary parts (bottom) of collisional modes in the case of $\bar{R}_0 = 10^{-5}$ (left) and $10^{-4}$ (right) as a function of wavenumber $k$.
    Red and magenta (blue and teal) colors correspond to the near/gapless (far/gapped) modes named on isotropic and homogeneous background.
    Grey dotted lines in the upper panels illustrate the avoidance boundaries defined by $\abs{\mathrm{Re}\,\omega} = \abs{k}$, within which the angular integrals could be singular for real waves.
    Green dash-dot lines are at $k = -\Phi_z$ with $K=0$ and $\mathrm{Re}\,\omega = -\Phi_0$ with $\mathrm{Re}\,\Omega = 0$.
    In the right panels, the damping near/gapless modes could not be reliably identified (see Appendix\,\ref{Sec:App} for details).}
    \label{fig:DR_branch}
\end{figure*}

\subsection{Dispersion relation \& collisional instability}\label{Sec:3.2_mode_analysis}
Figure\,\ref{fig:DR_branch} shows the dispersion relations for two reaction-rate configurations, $\bar{R}_0 = 10^{-5}$ (left) and $10^{-4}$ (right panels).
In both cases, multiple growing modes appear, but their dominant instabilities exhibit distinct wavenumber dependences.
For the smaller anti-neutrino reaction rate, the near/gapless modes (red and magenta) develop with a sharply localized maximum growth rate in the vicinity of the zero mode ($k=0$).
In contrast, for the larger case, the far/gapped branches (blue and teal) become dominant and achieve their maximum growth rate at finite wavenumbers.
We first characterize these different unstable branches and then explore the physical origin of their distinct behaviors.

Note that the dispersion relations are shifted by the lepton current $\Phi^{\mu}$ into the $\omega-k$ plane.
In the isotropic neutrino background, since the spatial phase shift becomes zero, the zero mode $k=0$ is identical to the physical homogeneous mode $K=0$.
However, in the present anisotropic environments, the coincidence is broken.
Green dash-dot lines correspond to a spatially homogeneous mode $K=0$ satisfying $k = -\Phi_z$ and a stationary mode $\Omega = 0$ satisfying $\omega = -\Phi_0$, respectively.
In our setup, $\Phi_0 = 0.2$ and $\Phi_z \sim 0.06$ in both cases.

\subsubsection{Unstable branches}
For the smaller reaction rate case $\bar{R}_0 = 10^{-5}$, the near/gapless branches dominate the flavor instabilities.
In particular, the Near 1 branch (red) exhibits a sharp maximum growth rate with $\mathrm{Im}\,\omega \sim 9.0\times 10^{-4}$ around $k=0$, as shown in the lower-left panel of Fig.\,\ref{fig:DR_branch}.
The peak is not exactly localized at the zero mode but at $(\mathrm{Re}\,\omega, k) \simeq (3.7\times 10^{-4}, 4.3\times 10^{-4})$.
The nonzero real frequency is comparable to the collision scale rather than the collective scale \cite{Fiorillo:2026c}\footnote{
In that sense, the near/gapless mode is strictly a collision-scaled gapped branch.}.
The branch extends over a broad range of finite wavenumbers and then asymptotically follows the near-luminal relation,
\begin{equation}
    \omega(k) \simeq v_{\mathrm{ph}}k + \ii\gamma(k),
\end{equation}
with a phase velocity $v_{\mathrm{ph}} \simeq 1$ and imaginary part $\gamma$.
Meanwhile, two far/gapped branches (blue and teal) with the real frequencies on the collective scale, $\mathrm{Re}\,\omega = \mathcal{O}(\mu)$, are damping modes in this configuration.

In contrast, for the larger case $\bar{R}_0 = 10^{-4}$, the flavor instability is instead dominated by the two far/gapped branches, whose real frequencies remain on the collective scale rather than the collision scale.
Unlike the near/gapless case, their maximum growth rates appear at finite wavenumbers, forming relatively broad peaks in the lower-right panel.
The corresponding real frequencies approach the near-luminal regime, $\abs{\mathrm{Re}\,\omega} \simeq \abs{k}$, around the maximum growth rates, whereas towards the zero mode $k=0$, the branches retain the finite real frequencies with collective scale and enter the super-luminal regime, $\abs{\mathrm{Re}\,\omega} > \abs{k}$.
We note that the near/gapless branches, which are expected to be damped in this case, could not be reliably identified due to numerical difficulties associated with collisional spurious modes, as discussed in Appendix\,\ref{Sec:App}.

\subsubsection{Collisional corrections for Far/Gapped modes}
To understand the origin of the far/gapped modes, we examine the impacts of collisions on the underlying collective modes.
In particular, although reaction rates are different between the left and right panels of Fig.\,\ref{fig:DR_branch}, the real parts of the eigenfrequencies follow nearly the same track.
The behaviors can be predicted by extending the discussion in Ref.\,\cite{Fiorillo:2026c}.
Since their real wave indices ($\mathrm{Re}\,\omega,k$) remain on the collective scale even in the collisionless limit, the collisional effects can be treated perturbatively when the reaction rates are sufficiently small.

A far/gapped branch possesses a real eigenfrequency $\omega_0 = \mathcal{O}(\mu)$ in the collisionless limit.
When collisions are taken into account, the correction to the complex eigenfrequency of the collective modes induces a flavor instability, growing or damping.
For sufficiently small reaction rates, the eigenfrequency can be expanded as
\begin{equation}
    \omega(k) \simeq \omega_0(k) + \delta\omega_{\mathrm{col}}(k),
\end{equation}
and the growth rate is given by
\begin{equation}
    \gamma(k) = \mathrm{Im}\,\delta\omega_{\mathrm{col}}(k).
\end{equation}
Focusing on the integral parts Eq.\,\eqref{eq:I_n_DR} in the dispersion relation, the gapped condition $\abs{\omega_0} \gg R_{\mathrm{EA}},\delta\omega_{\mathrm{col}}$ at the finite wavenumber $k=\mathcal{O}(\mu)$ can expand the denominator into
\begin{align}
    &\frac{1}{\omega_0+\delta\omega_{\mathrm{col}}-kv+\ii R_{\mathrm{EA}}} \notag \\
    &\qquad= \frac{1}{\omega_0-kv} - \frac{\delta\omega_{\mathrm{col}} + \ii R_{\mathrm{EA}}}{(\omega_0-kv)^2} + \mathcal{O}(R_{\mathrm{EA}}^2),
\end{align}
where we impose $\delta\omega_{\mathrm{col}} = \mathcal{O}(R_{\mathrm{EA}})$ outside the resonancelike regime where the growth rate can instead scale as $\sqrt{\mu R}$.
This expansion is valid provided that $\abs{\delta\omega_{\mathrm{col}}+\ii R_{\mathrm{EA}}} \ll \abs{\omega_0 - kv}$ over the relevant phase-space domain.
The integral part is then decomposed into
\begin{align}
    I_n &= I_n(\omega_0 + \delta\omega_{\mathrm{col}}) \notag \\
        &= \int\dd\Gamma\, G_{ex}\left[\frac{v^n}{\omega_0-kv}-\frac{\left(\delta\omega_{\mathrm{col}} + \ii R_{\mathrm{EA}}\right) v^n}{(\omega_0-kv)^2}\right] + \mathcal{O}\left(R_{\mathrm{EA}}^2\right) \notag \\
        &= I_n(\omega_0) + \frac{\partial I_n}{\partial \omega}\Big|_{\omega_0}\delta\omega_{\mathrm{col}} - \ii\delta I_n^{\mathrm{EA}}(\omega_0) + \mathcal{O}\left(R_{\mathrm{EA}}^2\right),
\end{align}
where the first term is the collisionless part, and the second and third corrections give rise to an unstable branch.
Note that the third correction can be regarded as the derivative term in terms of $R_{\mathrm{EA}}$.
This expansion is applicable to Eq.\,\eqref{eq:determinant_D}, and then the dispersion formula is similarly approximated as
\begin{align}
    0 &= \mathcal{D}(\omega_0+\delta\omega_{\mathrm{col}}) \notag \\
        &= \mathcal{D}(\omega_0) + \frac{\partial\mathcal{D}}{\partial\omega}\Big|_{\omega_0}\delta\omega_{\mathrm{col}} - \ii C(\omega_0) + \mathcal{O}(R_{\mathrm{EA}}^2).
\end{align}
Since the first term vanishes in the collisionless limit, $\omega=\omega_0$, the correction to the root is obtained by
\begin{equation}
    \delta\omega_{\mathrm{col}} = \ii \frac{C(\omega_0)}{\mathcal{D}^{\prime}(\omega_0)} + \mathcal{O}(R_{\mathrm{EA}}^2).
\end{equation}
The correction $\delta\omega_{\mathrm{col}}$ is purely imaginary because both the derivative term $\mathcal{D}^{\prime}(\omega_0)$ and the collisional correction $C(\omega_0)$ are exactly real for the real eigenfrequency $\omega_0$.
It is consistent with the previous analytical (approximated) solutions for the isotropic, homogeneous, and monochromatic neutrino gases, in which collision term is contained only in the imaginary parts and does not change the real ones \cite{Lin:2023,Liu:2023,Xiong:2023b}.
This is why, provided the neutrino distribution remains unchanged, the real part of the unstable branch appears the same for both reaction rate settings in the upper panels of Fig.\,\ref{fig:DR_branch}.

The collisional correction also provides insight into the wavenumber dependence of the growth rate.
In particular, we focus on the near-luminal regime to understand why the far/gapped branches achieve their maximum growth rates at finite wavenumbers.

In the super-luminal regime $k\simeq 0$, since the denominator becomes independent of velocity, the correction contributes a flavor instability only through the angular moments $\int\dd v\,G_{ex}v^n$.
On the other hand, in the near-luminal regime $k\simeq \mathrm{Re}\,\omega$, the angular dependence in the denominator enhances a forward-peaked contribution:
\begin{equation}
    \mathrm{Re}\,\omega - kv = k(v_{\mathrm{ph}}-v) \simeq k(1-v).
    \label{eq:near_luminal}
\end{equation}
Consequently, the contribution in the collisional correction $\delta I_n^{\mathrm{EA}}$ concentrates on the spatially-inhomogeneous mode $k\neq 0$ through the intrinsic forward-peaked angular distributions and potentially leads to a maximal growing mode at a finite wavenumber $k$.
This provides a possible explanation for the peak of the far/gapped branch appearing far from the origin $(\mathrm{Re}\,\omega,k) = (0,0)$.
If the neutrino background now considered has a backward-peaked angular distribution, the unstable branch would peak at $k \simeq - \mathrm{Re}\,\omega$ with a negative phase velocity.

\begin{figure*}
    \centering
    \subfloat[Near 1 mode with $\bar{R}_0 = 10^{-5}$]{
        \includegraphics[width=0.49\linewidth]{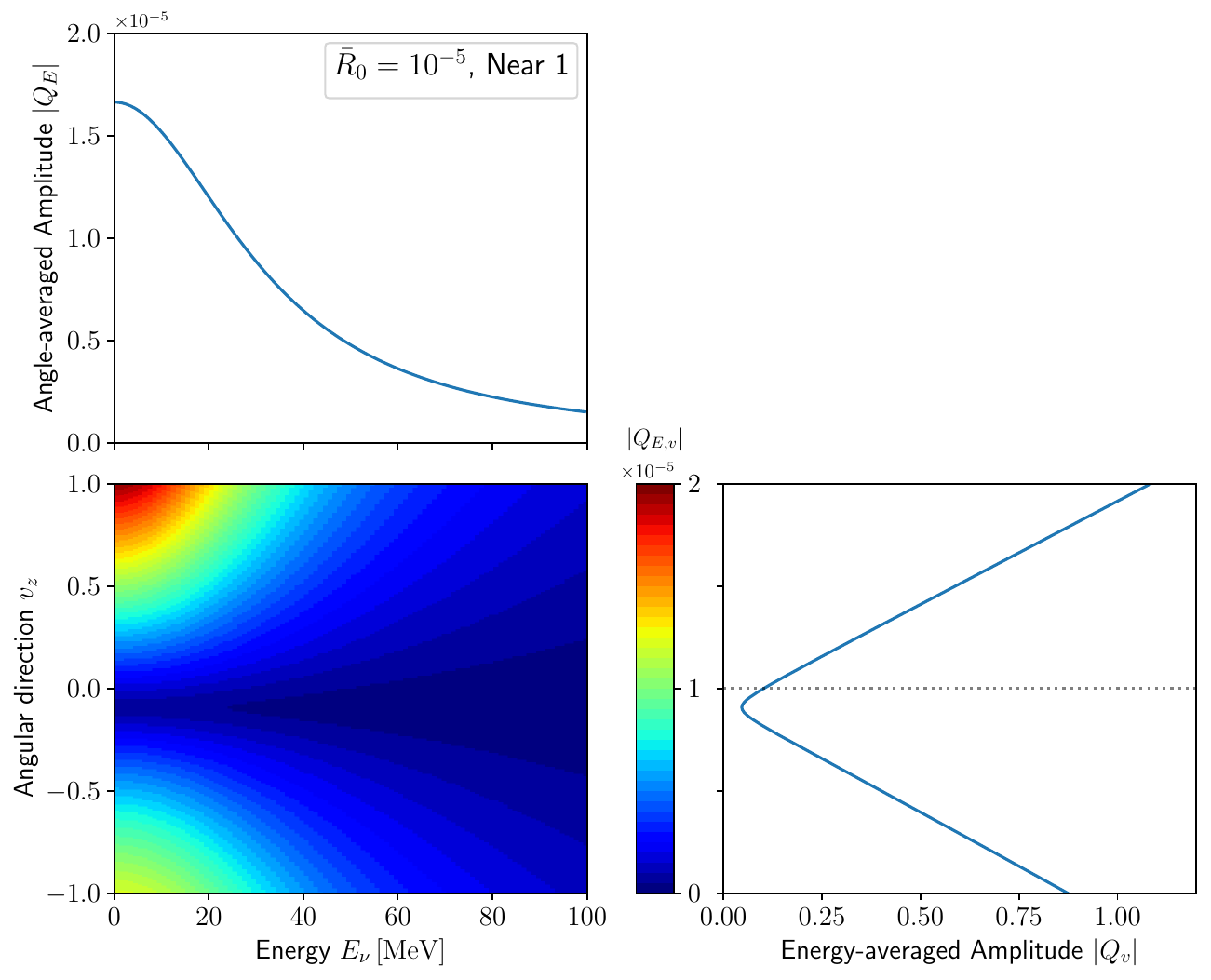}}
    \subfloat[Near 2 mode with $\bar{R}_0 = 10^{-5}$]{
        \includegraphics[width=0.49\linewidth]{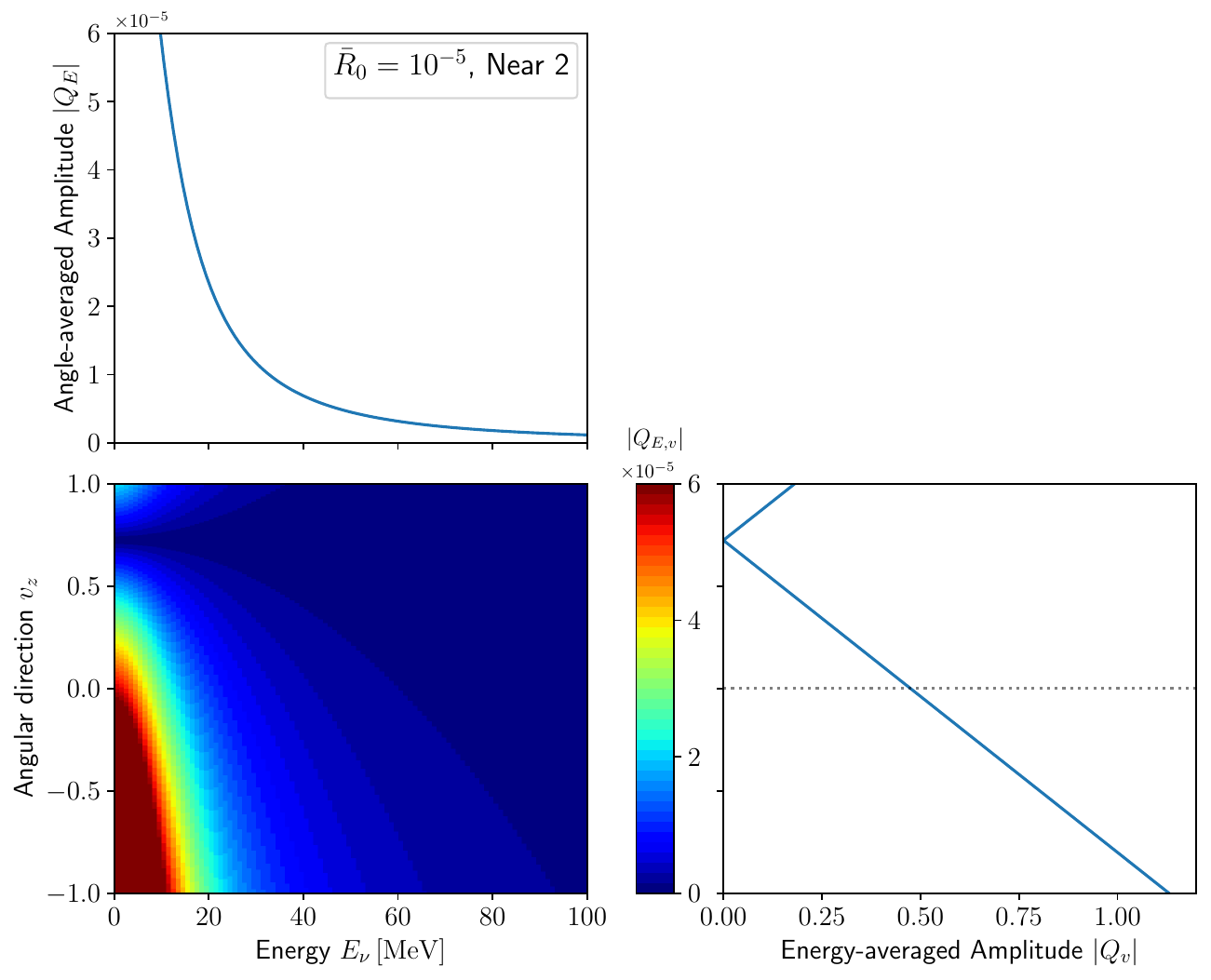}}
    \par\medskip
    \centering
    \subfloat[Far 1 mode with $\bar{R}_0 = 10^{-4}$]{
        \includegraphics[width=0.49\linewidth]{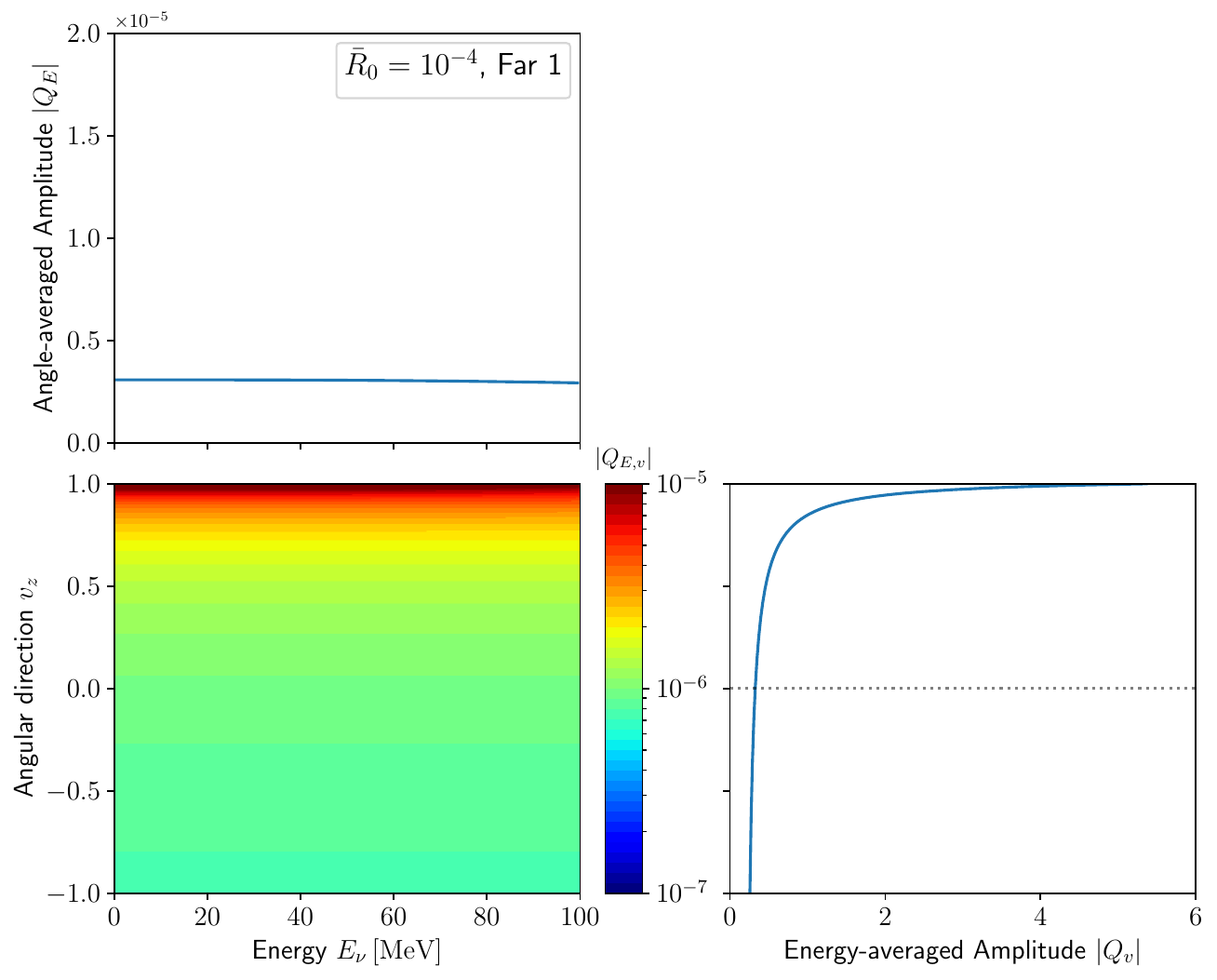}}
    \subfloat[Far 2 mode with $\bar{R}_0 = 10^{-4}$]{
        \includegraphics[width=0.49\linewidth]{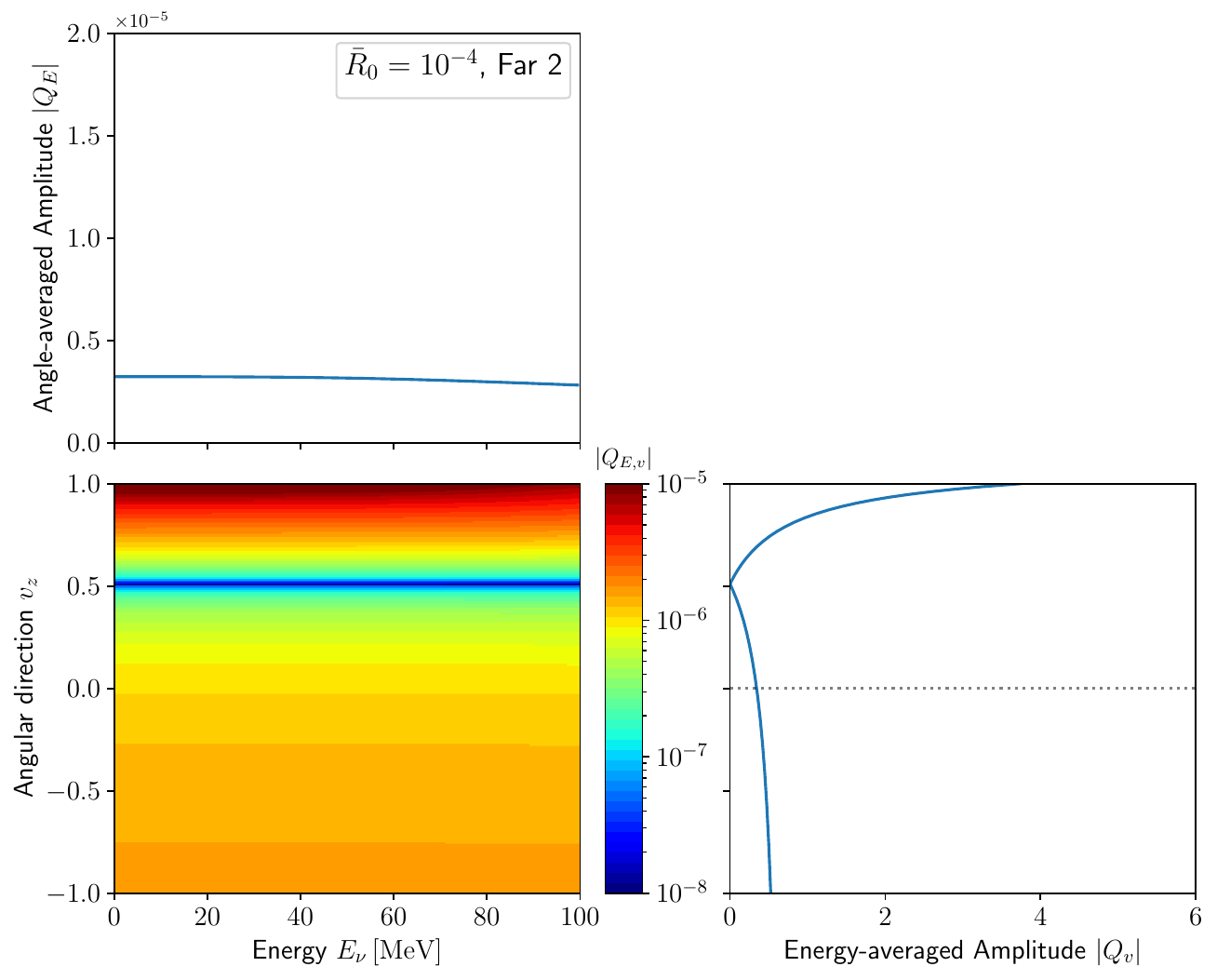}}
    \caption{Eigenvectors $\abs{\tilde{Q}_{E_{\nu},v,\omega,k}}$ without intrinsic neutrino distribution $G_{ex}(E_{\nu},v)$ at the peak growing modes in the case of $\bar{R}_0 = 10^{-5}$ (top) and $10^{-4}$ (bottom).
    Upper left and bottom right panels in each sub-figure correspond to the amplitude averaged over angle and energy distributions.}
    \label{fig:DR_eigen}
\end{figure*}

\subsubsection{Angular contributions to Near/Gapless modes}
The near-luminal enhancement discussed above may also appear in the near/gapless branch.
Indeed, the Near 1 branch in Fig.\,\ref{fig:DR_branch} exhibits a small bump around $k\sim -0.4$ and a plateau around $k\sim 0.1$, where the phase velocity is close to unity and the far/gapped branches have damping peaks.
These features may reflect the enhanced contributions from the forward angular directions, similar to the mechanism discussed for the far/gapped branch around Eq.\,\eqref{eq:near_luminal}.

Nevertheless, the near/gapless branch exhibits the maximum growth rate in the vicinity of the zero mode.
In this region, the eigenfrequency and wavenumber are both comparable to the collision scale, so that the denominator of $I_n$ remains small over a broad region of angular directions, rather than being restricted to the forward directions as in Eq.\,\eqref{eq:near_luminal}.
Consequently, the collision-scaled (or negligibly small) wavenumber implies that contributions from not only forward-going neutrinos but also a broad range of angular directions can be enhanced.
This broad angular contribution may account for the much stronger enhancement of the growth rate near the zero mode compared to the small bump at finite wavenumbers.

\subsection{Eigenvector structures}\label{Sec:3_eigenvec}
The dispersion relations characterize the unstable branches through their complex eigenfrequencies.
Meanwhile, the corresponding eigenvectors reveal how the unstable modes develop flavor coherence in the phase space.
Before proceeding to nonlinear flavor evolution, we investigate their energy and angular dependences for each unstable branch.

Figure\,\ref{fig:DR_eigen} displays the amplitude of the eigenvector in the ($E_{\nu},\,v$) plane for the maximum growing mode of each unstable branch.
A clear distinction between the near/gapless and far/gapped modes lies in their energy dependence, arising from the collision term in the denominator of Eq.\,\eqref{eq:eigenvec}.

In the case of the near/gapless modes provided with the smaller reaction rate, the wave indices $(\omega,k)$ in the denominator are scaled by reaction rates $R_{\mathrm{EA}}$.
Thereby, the eigenvector has an energy structure of $\propto 1/(1+E_{\nu}^2)$.
In contrast, the far/gapped branch is characterized by collective scale $\mu$, and the peak growing mode is also far from the zero mode $k=0$.
This means that the real frequency and wavenumber are much larger than reaction rate $R_{\mathrm{EA}}$ and the growth rate, and the eigenvectors are then approximated as
\begin{equation}
    \tilde{Q}_{E_{\nu},v} \propto \frac{\tilde{a}_0-\tilde{a}_z v}{\mathrm{Re}\,\omega - kv},
\end{equation}
which is insensitive to neutrino energy and depends only on angles.
Such energy dependence for each unstable branch has been reported by our previous study \cite{Zaizen:2025} with multi-energy but isotropic and homogeneous neutrino gases.

Another distinction lies in their angular dependences and requires evaluating both the numerator and denominator.
The general eigenvector could have a sharp enhancement coming from the denominator:
\begin{equation}
    \tilde{Q} \propto \frac{\tilde{a}_0 - \tilde{a}_z v}{k\left(v_{\mathrm{ph}}-v\right) + \ii\left(\mathrm{Im}\,\omega + R_{\mathrm{EA}}\right)}.
\end{equation}
Particularly, considering the negligible imaginary parts, it becomes singular at $v = v_{\mathrm{ph}}$.
In other words, the far/gapped branch has a sharp peak in the eigenvector at $v = v_{\mathrm{ph}}$.
Since the dispersion relation appears within the near-luminal and super-luminal regimes according to our numerical demonstrations in Fig.\,\ref{fig:DR_branch}, its eigenvector likely becomes strongly peaked in the forward angular directions, as shown in Fig.\,\ref{fig:DR_eigen}.

In the near/gapless limit, $\omega,k = \mathcal{O}(R_{\mathrm{EA}})$, all terms in the denominator remain comparable.
Nevertheless, the angular structure can be roughly simplified into
\begin{equation}
    \tilde{Q}_v \propto \frac{\tilde{a}_0-\tilde{a}_zv}{A-kv}
\end{equation}
with a constant $A$ containing the other properties.
According to our numerical analysis mentioned above, $\abs{A} \gtrsim \abs{k}$ is satisfied around the peak, and then the eigenvector becomes approximately linear in $v$ with a minimum near $v = \mathrm{Re}\,\tilde{r}$.
Recalling the discussion about Eq.\,\eqref{eq:root_eigenvec}, the eigenvector further could have a slight offset, which is actually observed in the upper-left panel of Fig.\,\ref{fig:DR_eigen}.

\begin{figure*}
    \centering
    \includegraphics[width=1.\linewidth]{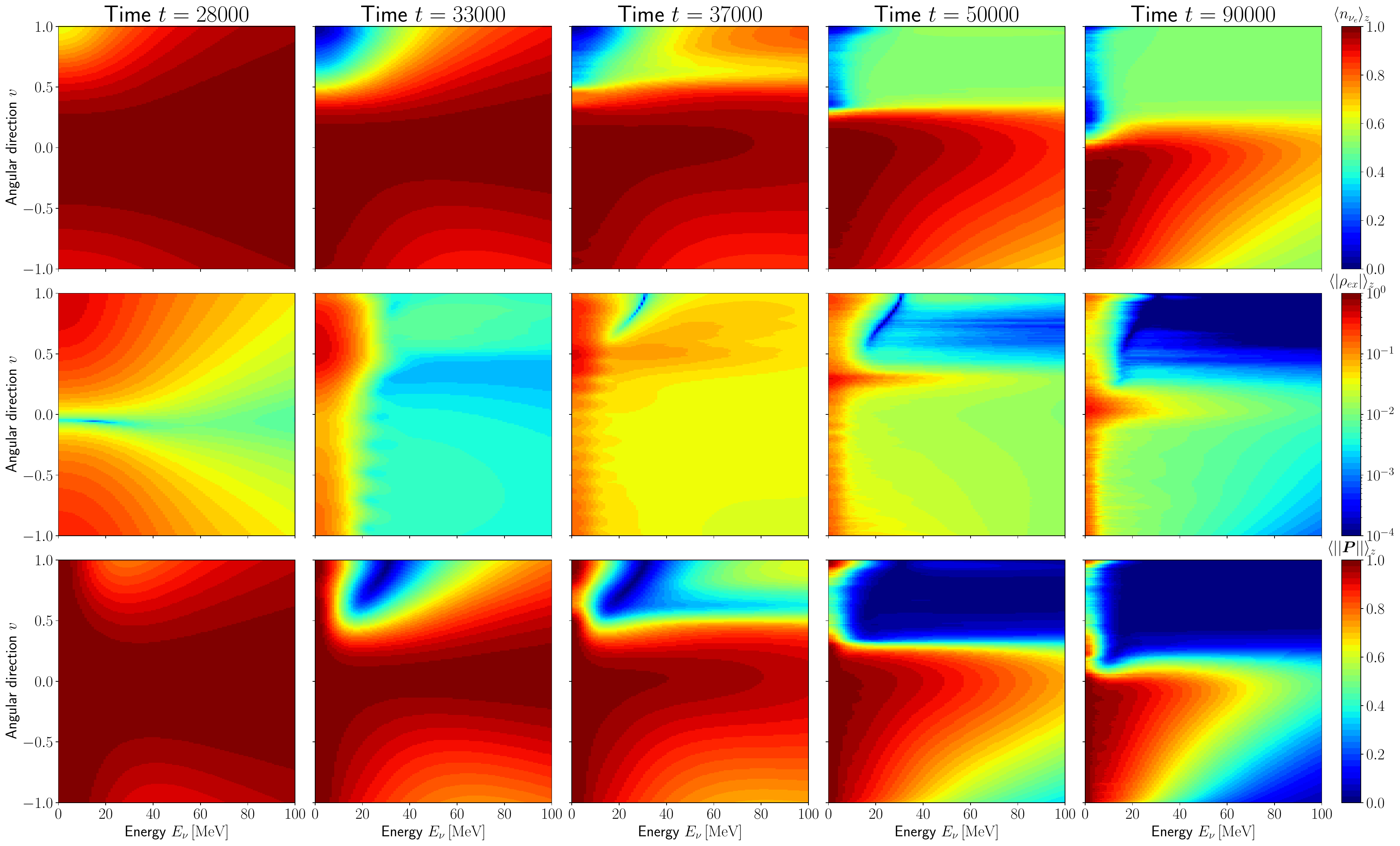}
    \caption{Time evolution of the number density $\langle n_{\nu_e}\rangle_z$ (top), the flavor coherence $\langle \abs{\rho_{ex}}\rangle_z$ (middle), and the norm $\langle \norm{\boldsymbol{P}}\rangle_z$ (bottom) of the polarization vector averaged over space in the case of $\bar{R}_0 = 10^{-5}$ where the near/gapless mode is dominant.
    See Ref.\,\cite{Zaizen:2026d} for movies showing the full flavor evolution corresponding to these snapshots.
    }
    \label{fig:n_ev_R01}
\end{figure*}

\section{Nonlinear Flavor Evolution}\label{Sec:4_Nonlinear}
\begin{figure}[t]
    \centering
    \includegraphics[width=1.\linewidth]{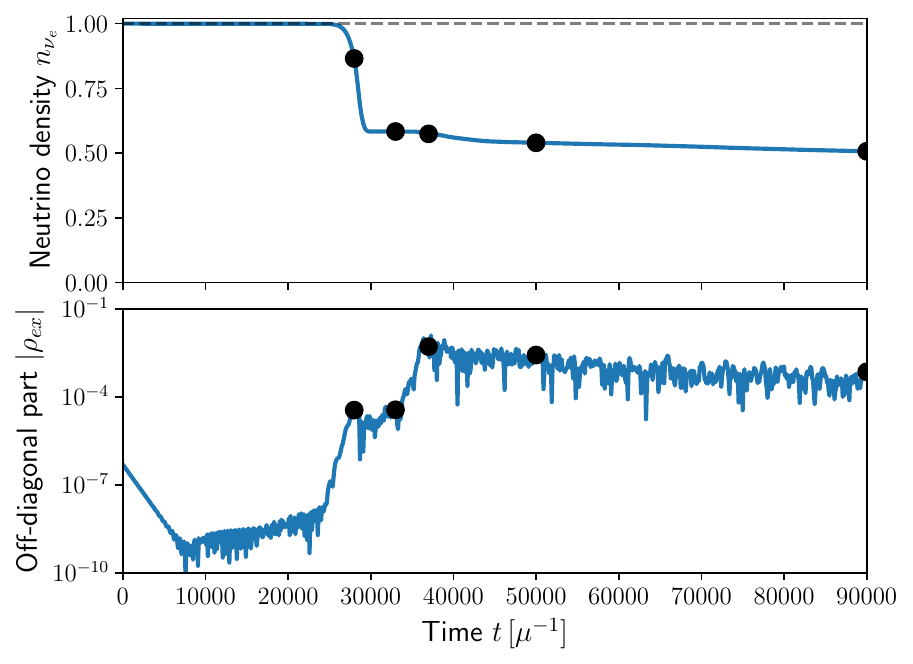}
    \caption{Time evolution of number density (top) and flavor coherence (bottom) averaged over phase space.
    Black dots correspond to the time snapshots in Fig.\,\ref{fig:n_ev_R01}.}
    \label{fig:num_dens_R01}
\end{figure}
In this section, we perform the nonlinear calculation of CFC on multi-energy, multi-angle, and spatially-inhomogeneous neutrino background in both cases of reaction rates and directly compare it to the eigenmode analysis in the previous section.
To describe the impact of collisional decoherence obviously, we discuss the flavor evolution within the polarization vector configuration, not directly using the density matrix;
\begin{equation}
    \rho = \frac{\mathrm{Tr}\rho}{2}\mathbbm{I}_2 + \frac{1}{2}\boldsymbol{P}\cdot\boldsymbol{\sigma},
\end{equation}
where $\boldsymbol{P}$ denotes the polarization vector and the trace part is hereafter called $P_0$, which is composed of the total flavor population.
In neglecting the diagonal collision term, the quantity does not change during the flavor evolution.
Thereby, normalizing the polarization vector $\boldsymbol{P}$ by $P_0$ separates the flavor polarization from the overall occupation $f_{\nu_{\alpha}}$.

\subsection{Near/Gapless mode dominant case}
First, we focus on the case of $\bar{R}_0 = 10^{-5}$ for which the near/gapless branch dominates the instability, which was shown to bring a flavor swap in our previous study \cite{Zaizen:2025} in the isotropic and homogeneous background.
Figure\,\ref{fig:n_ev_R01} shows the time evolution of the spatially-averaged $\nu_e$ population (top)
\begin{equation}
    \langle n_{\nu_e}\rangle_z = \frac{\int\dd z\, \left(P_0+P_3\right)/2}{\int\dd z\, P_0}
\end{equation}
spatially-averaged flavor coherence (middle)
\begin{equation}
    \langle \abs{\rho_{ex}}\rangle_z = \frac{\int\dd z\, \sqrt{P_1^2+P_2^2}\Big/2}{\int\dd z\, P_0},
\end{equation}
and the spatially-averaged norm of the polarization vector (bottom)
\begin{equation}
    \langle \norm{\boldsymbol{P}}\rangle_z = \frac{\int\dd z\, \norm{\boldsymbol{P}}}{\int\dd z\, P_0}
\end{equation}
from the left to the right panels with time, respectively.

At the early epoch $t=28000$, the flavor coherence demonstrates the same structure as the eigenvector in the upper left panels of Fig.\,\ref{fig:DR_eigen}.
The subsequent flavor evolution shows a flavor swap, $n_{\nu_e} \sim 0$, in lower energy and forward angular direction sides (upper left regions in the panel) at $t=33000$.
After that, the flavor coherence significantly decreases in the other regions on the phase space, particularly, neutrinos near the flavor-swap region are completely decohered, and the corresponding norm approaches zero (in the bottom panel).

Since the collisional damping is proportional to the instantaneous flavor coherence $\propto - R_{\mathrm{EA}}\rho_{ex}$, the flavor states near the equipartition where flavor coherence is maximally growing are predominantly collapsed.
The energy structure is understood from the eigenvector and the decoherence rate.
As mentioned in Sec.\,\ref{Sec:3_eigenvec}, the energy dependence of the eigenvector is determined by the nature of the near/gapless mode.
The decoherence rate for the norm of the polarization vector is then given by
\begin{equation}
    \dd_t\norm{P}^2 \propto - R_{\mathrm{EA}}\abs{\rho_{ex}}^2 \propto - \frac{R_{\mathrm{EA}}}{1 + R_{\mathrm{EA}}^2},
\end{equation}
so that the shrink in the norm is greatest at the intermediate energies rather than at the lowest.
Similarly, regions with relatively large eigenvector amplitudes at lower energy and backward direction side (lower left region) are damped before substantial flavor conversion develops.

In the middle epoch at $t=37000$, the flavor coherence is restored even though neutrinos have undergone decoherence due to collisions.
This means that the neutrino ensemble is in a metastable state with respect to flavor conversion.
The behaviors are clarified in the bottom panel of Fig.\,\ref{fig:num_dens_R01}, in which flavor coherence averaged over the whole phase space $(z,E_{\nu},v)$ is shown, and black dots correspond to each time snapshot shown in Fig.\,\ref{fig:n_ev_R01}.
It is obvious that the flavor coherence experiences the secondary exponential growth and reaches a linear saturation again.
The secondary flavor evolution has been observed in some previous studies on isotropic and homogeneous neutrino gases \cite{Johns:2023c,Froustey:2025,Ng:2026}.

\begin{figure}[t]
    \centering
    \includegraphics[width=1.\linewidth]{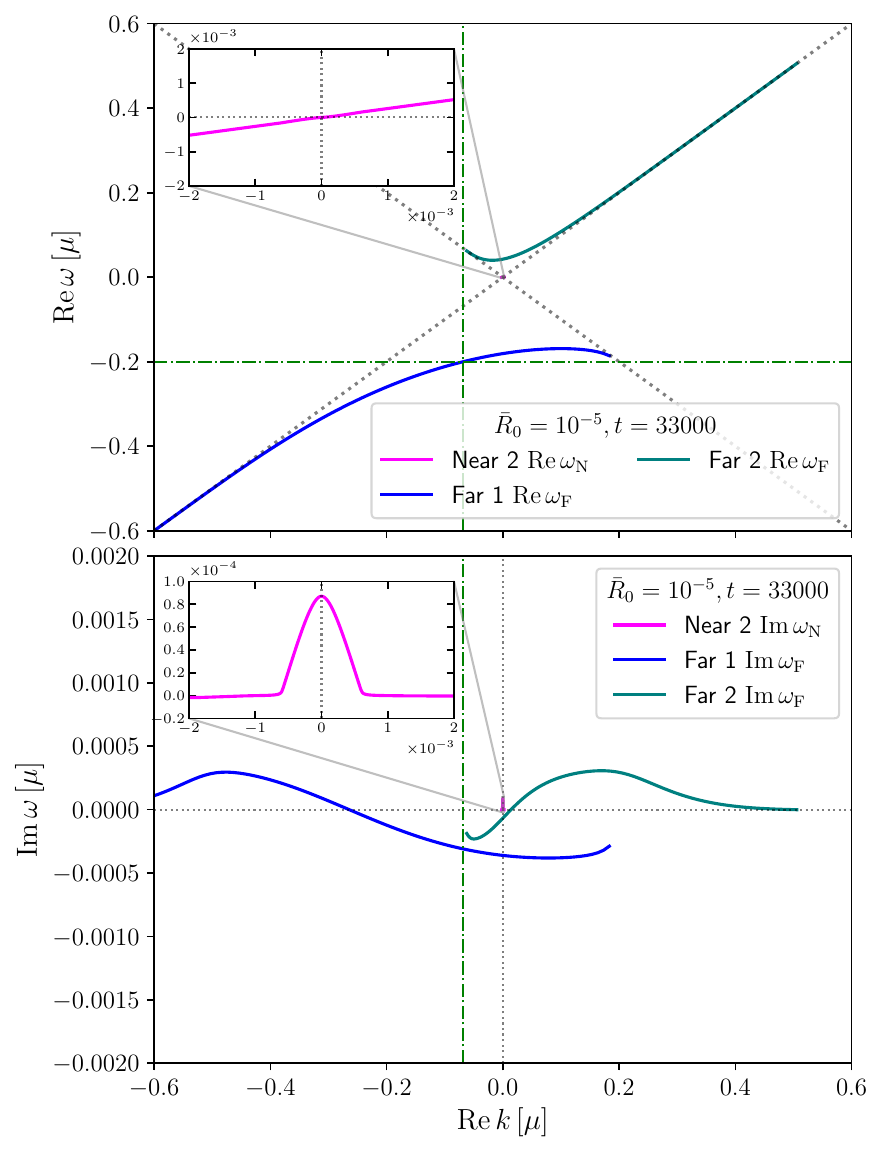}
    \caption{Same as Fig.\,\ref{fig:DR_branch}, but the dispersion relation for coarse-grained neutrino distribution of the post-CFC flavor state at $t = 33000$ in the case of $\bar{R}_0 = 10^{-5}$.}
    \label{fig:post_CFC}
\end{figure}
Focusing on the instantaneous neutrino flavor states at $t=33000$ in Fig.\,\ref{fig:n_ev_R01}, the flavor coherence is relatively small, particularly in the high-energy side, while in the low-energy side, the flavor coherence still survives but is also sub-dominant.
Therefore, we regard the coarse-grained neutrino ensembles as flavor states within the linear phase and revisit the instantaneous dispersion relation.
Fig.\,\ref{fig:post_CFC} displays the dispersion relation for the post-CFC neutrino distributions at $t=33000$.
It clarifies that the two far/gapped branches change from damping to growing.
Since the phase-space structure of the eigenvector is determined mainly by the real part of $(\omega,k)$ and the real part is almost the same as the case of $\bar{R}_0 = 10^{-4}$, the shape should be similar to ones shown in the bottom panels of Fig.\,\ref{fig:DR_eigen}.
This means that the secondary growth should exhibit relatively weak energy dependence and also tend to appear strongly in the forward directions.
Furthermore, the seed perturbation itself for the secondary growth is already anisotropic, so that the distinction between the forward and backward directions can be enhanced.
The behaviors can be confirmed from the flavor coherence at $t=37000$ in Fig.\,\ref{fig:n_ev_R01}.
Note that the variation in number density after the secondary flavor conversion looks minor in the top panel of Fig.\,\ref{fig:num_dens_R01}.
This is because a substantial fraction of the neutrino population has already undergone flavor swap during the initial conversion.

The matured flavor coherence is decohered again due to the incoherent collisions, and the norm shrinks in the broader energy range in the late epoch at $t=50000$ and $90000$.
Since the third component (flavor eigenstate) of the polarization vector is almost identical to the norm after sufficiently undergoing collisional decoherence, the survival probability is determined simply as \cite{Zaizen:2025}
\begin{equation}
    P_{ee} = \frac{1}{2}\left(P_0 + P_3\right)/P_0 \simeq \frac{1}{2}\left(1 + \norm{\boldsymbol{P}}/P_0\right).
\end{equation}
This equation says that completely collapsed polarization vector corresponds to flavor equipartition.
As a consequence, the neutrino ensemble approaches a flavor equipartition state in the asymptotic state.

\begin{figure}[t]
    \centering
    \includegraphics[width=1.\linewidth]{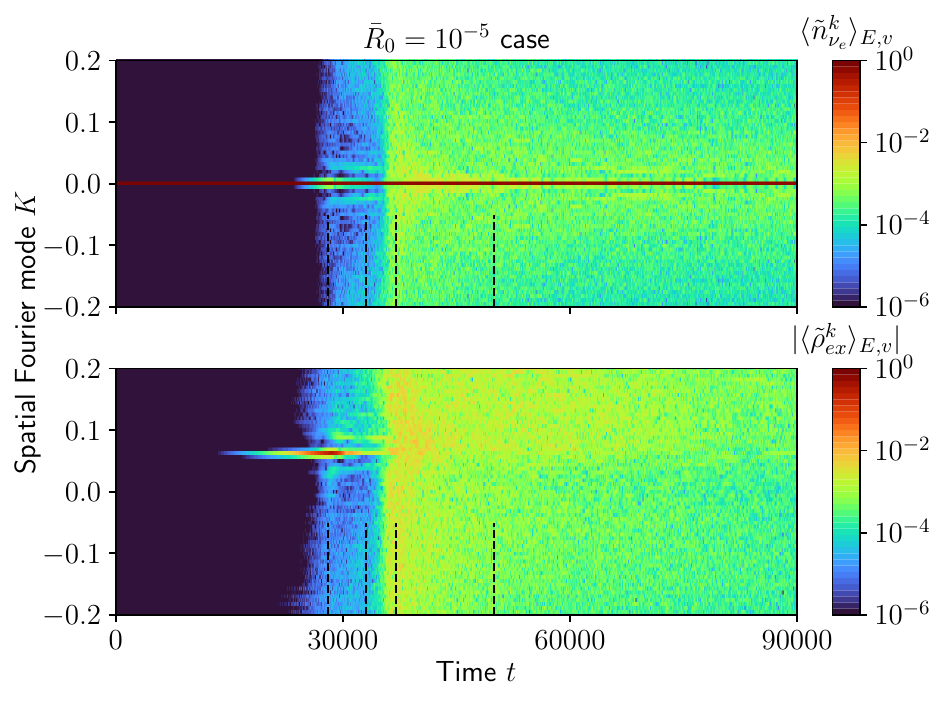}
    \caption{Time evolution of spatial Fourier mode $K$ in number density (top) and flavor coherence (bottom) averaged over momentum space $(E_{\nu},v)$.
    Vertical black dashed lines correspond to the time snapshots in Fig.\,\ref{fig:n_ev_R01}.}
    \label{fig:Fourier_tk_R01}
\end{figure}
We have so far focused on the spatially-averaged quantities during the flavor evolution.
Figure\,\ref{fig:Fourier_tk_R01} displays the time evolution of the spatial Fourier modes $K$ in time.
At the early phase, flavor coherence in the bottom panel undergoes exponential growth only at a specific spatial mode $K\sim 0.06$, which is consistent with the dispersion relation shown in the left panel of Fig.\,\ref{fig:DR_branch}.
Note that the spatial variation is measured on the physical frame $(\Omega,K)$, not on the shifted frame $(\omega,k)$.

\begin{figure*}
    \centering
    \includegraphics[width=1.\linewidth]{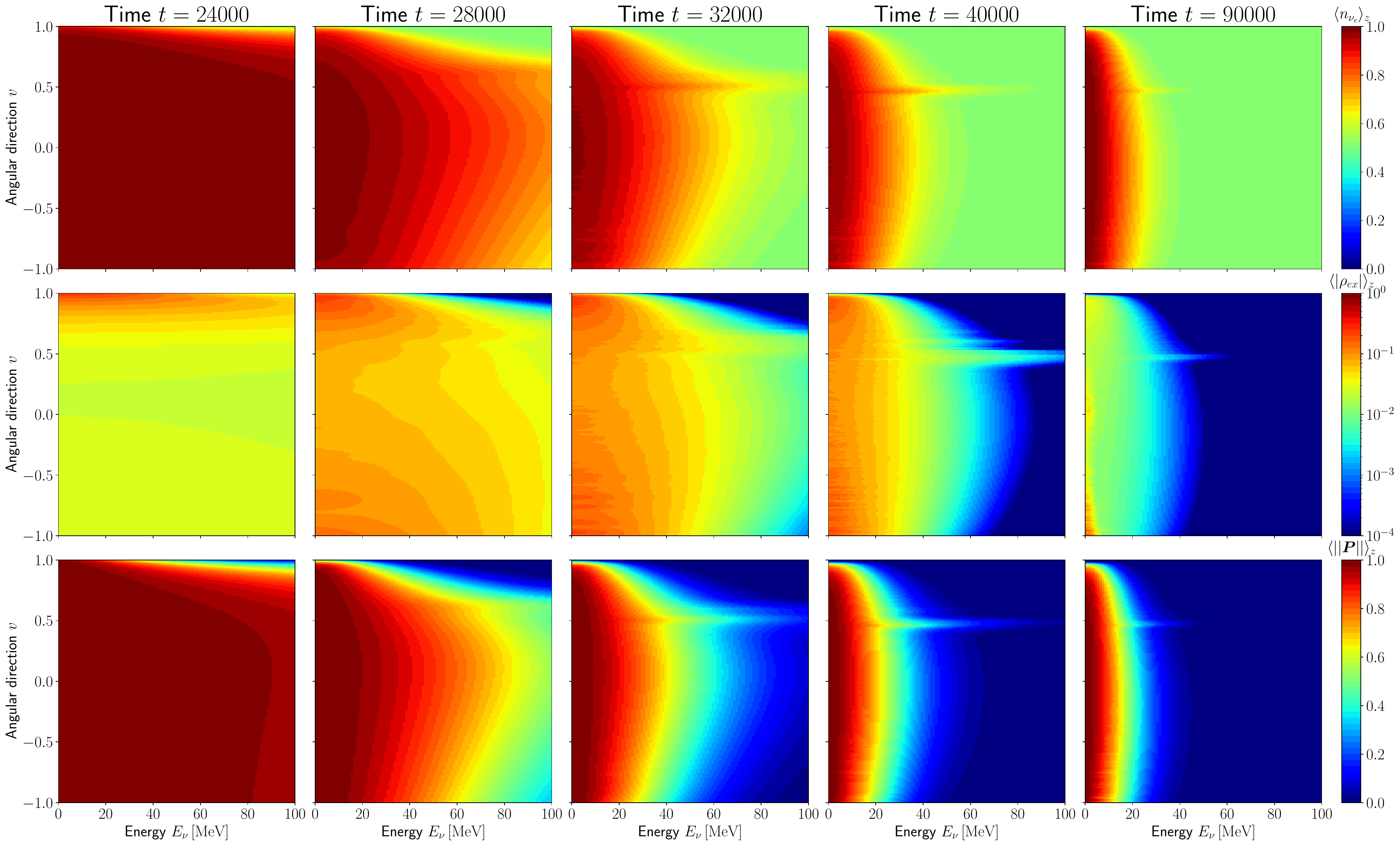}
    \caption{Same as Fig.\,\ref{fig:n_ev_R01}, but in the case of $\bar{R}_0 = 10^{-4}$ where the far/gapped mode is dominant.}
    \label{fig:n_ev_R1}
\end{figure*}
\begin{figure}
    \centering
    \includegraphics[width=1.\linewidth]{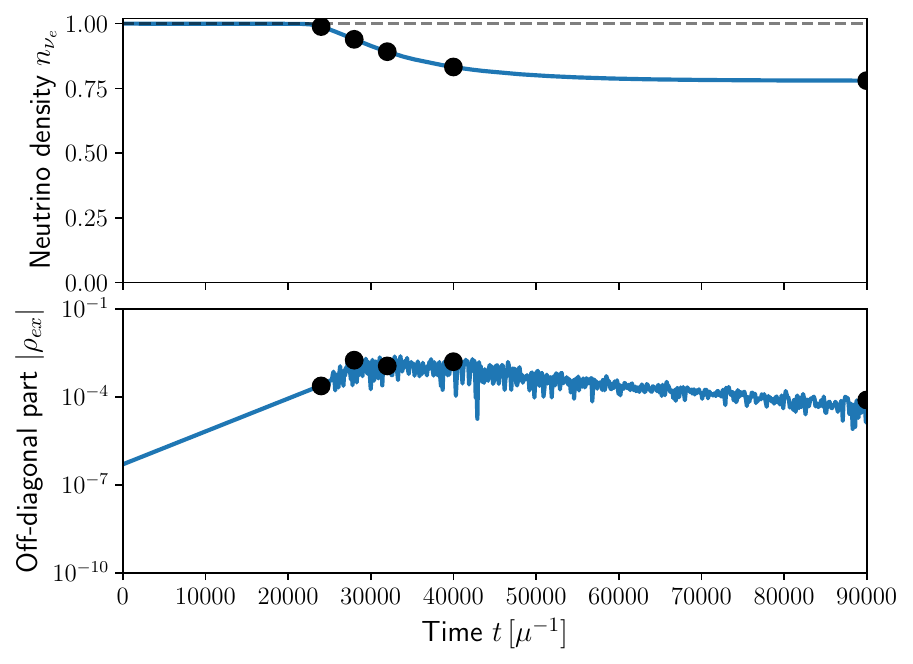}
    \caption{Same as Fig.\,\ref{fig:num_dens_R01}, but in the case of $\bar{R}_0 = 10^{-4}$.}
    \label{fig:num_dens_R1}
\end{figure}

After the linear saturation, finite flavor coherence starts to appear across broad spatial modes due to the wave-wave couplings.
Simultaneously, the prominently growing mode gradually decays due to collisional decoherence and has already been smeared by the onset of the secondary growth.
In the secondary exponential growth, flavor coherence widely grows but the value is not sufficiently large, as indicated by the instantaneous dispersion relation in Fig.\,\ref{fig:post_CFC}.
Therefore, the diagonal flavor population remains nearly spatially homogeneous, and the spatial inhomogeneity is minor as shown in the top panel.
This behavior is quite different from the development of a cascade in the FFC case \cite{Richers:2021,Zaizen:2021a,Wu:2021,Richers:2022,Bhattacharyya:2022,Zaizen:2023}.
In the case of collision-induced flavor conversion, collisional decoherence rather than flavor depolarization is a key ingredient in determining the asymptotic states.

\subsection{Far/Gapped mode dominant case}
Flavor evolution in the case of $\bar{R}_0 = 10^{-4}$ where two far/gapped modes are dominant is exhibited in Figs.\,\ref{fig:n_ev_R1} and \ref{fig:num_dens_R1}.
The eigenvectors are less dependent on neutrino energy, so that the evolution of flavor coherence looks strongly forwardly-peaked and almost iso-energetic at the early epoch $t=24000$.
Thereby, the high-energy components immediately experience heavier collisional decoherence, so that the norm starts to shrink.
As time passes, the other angular modes gradually grow and undergo a collisional decoherence similarly.
Consequently, neutrinos with energies above $20\,\mathrm{MeV}$ approach flavor equipartition states.

Away from the forward directions $v\simeq 1$, the asymptotic states are less sensitive to angular distributions.
However, the Far 2 mode brings a slight anisotropic structure with a minimum in its angular amplitude, different from the Far 1.
The result is the sharp structure around $v\sim 0.5$ where the amplitude of the eigenvectors is weaker than the other angular directions in considering the superposition of the Far 1 and 2 modes.

\begin{figure}[t]
    \centering
    \includegraphics[width=1.\linewidth]{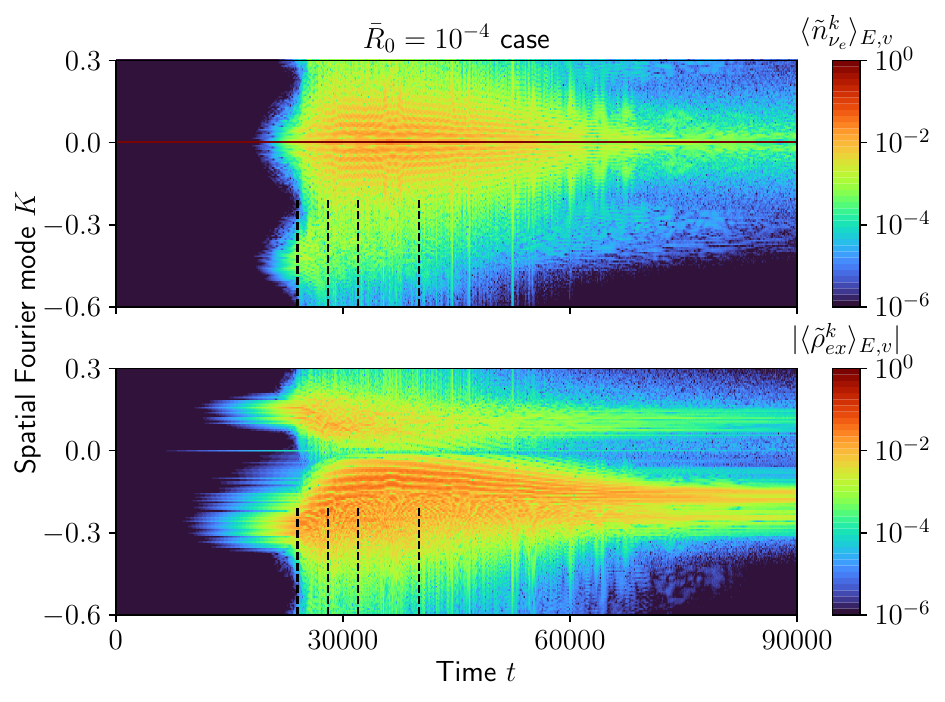}
    \caption{Same as Fig.\,\ref{fig:Fourier_tk_R01}, but in the case of $\bar{R}_0 = 10^{-4}$.}
    \label{fig:Fourier_tk_R1}
\end{figure}

The contributions of the multiple growing modes are also evident in the spatial evolution in Fig.\,\ref{fig:Fourier_tk_R1}.
The growth rates are comparable to the linear stability analysis in Fig.\,\ref{fig:DR_branch}, resulting in the bimodal structures in the spatial Fourier modes $K$ for flavor coherence.
Following the linear saturation, the mode spreading occurs similarly; however, this ceases without yielding sufficient spatial inhomogeneity due to collisional decoherence.
Unlike the near/gapless mode, the far/gapped mode attains its maximum growth rate at nonzero spatial wavenumbers $K\neq 0$.
Nevertheless, the resultant flavor conversion is likely to be homogeneous because the collisional decoherence is independent of the spatial structures and just damps the matured flavor coherence.
In other words, the resultant flavor conversion is largely determined by the norm of the polarization vector after collisional decoherence.

\section{Summary and Conclusions}\label{Sec:5_summary}
In this paper, we have investigated the linear stability and nonlinear flavor evolution of collision-induced neutrino-flavor conversion (CFC) in multi-energy, multi-angle, and spatially inhomogeneous neutrino gases.
We have focused on the non-resonancelike regime with forward-peaked angular distributions, extending our previous study of isotropic and homogeneous neutrino systems \cite{Zaizen:2025}.
By combining the dispersion relation approach with direct quantum kinetic simulations, we have examined how the properties of unstable collective modes are reflected in the subsequent flavor evolution and how collisional decoherence influences the asymptotic states.

Our linear stability analysis reveals distinct wavenumber dependences of the near/gapless and far/gapped modes.
The near/gapless branch exhibits a sharply localized peak in the growth rate near the zero mode $k=0$, whereas the far/gapped branches attain their maximum growth rates at finite wavenumbers on the collective scale.
The far/gapped branches exhibit nearly identical real parts of their eigenfrequencies for the two reaction-rate configurations.
This behavior can be understood from the perturbative analysis, which shows that the real parts are largely determined by the underlying collisionless collective modes, while the leading-order collisional corrections determine whether the collective modes are growing or damping.
The eigenvector corresponding to each branch exhibits distinct energy and angular dependences:
the near/gapless modes exhibit strong energy dependence with distinct angular structures, whereas the far/gapped modes are insensitive to energy and strongly enhanced in the forward angular directions.

These eigenvector structures are closely related to the nonlinear flavor evolution.
In the near/gapless-dominated case, flavor conversion initially develops preferentially in the low-energy and forward angular regions, where a prominent flavor swap occurs.
Collisional decoherence subsequently reduces the norm of the polarization vector over a broader region of the phase space, driving the remaining neutrinos towards flavor equipartition.
In contrast, the far/gapped-dominated case exhibits nearly energy-independent growth in flavor coherence, followed by heavier decoherence at higher energies and predominantly flavor equipartition.
The coexistence of two growing far/gapped modes with different angular structures also leaves a localized angular structure in the resulting flavor distribution.

An additional feature of the near/gapless-dominated case is the occurrence of secondary flavor growth after the initial flavor conversion and decoherence.
By revisiting the linear stability analysis using the instantaneous flavor distributions at the post-CFC states, we find that the far/gapped branches, which were initially damped, become unstable.
This demonstrates that the evolving neutrino background can change the stability of collective modes and trigger subsequent episodes of flavor conversion.

We have further examined the development of spatial inhomogeneities through the Fourier decomposition of the quantum kinetic simulations.
Although the intrinsic instabilities amplify flavor coherence on specific spatial wavenumbers, nonlinear mode coupling subsequently spreads it over a broader range of scales.
Nevertheless, collisional decoherence limits the persistence of flavor coherence and suppresses the development of strong spatial inhomogeneities in the flavor populations.
This behavior is observed even when the dominant unstable branches attain their maximum growth rates at finite wavenumbers.
In the presence of incoherent collisions, the competition between flavor depolarization and collisional decoherence determines the asymptotic flavor states.

Overall, our results demonstrate that the nonlinear outcome of collisional neutrino-flavor conversion cannot be understood from the maximum growth rate alone.
The energy and angular structures of the unstable eigenmodes provide essential information about where flavor conversion develops, while the evolution of the neutrino background and collisional decoherence determine how flavor conversion proceeds and which flavor states survive.
These findings provide a basis for understanding collisional flavor conversion beyond homogeneous and isotropic neutrino gases and motivate further investigations incorporating dynamically evolving backgrounds \cite{Xiong:2026} and more complete reaction processes \cite{Liu:2024,Ng:2026} in astrophysical environments.

\section{Acknowledgments}
We are grateful to Jiabao Liu, Chinami Kato, Shota Takahashi, Hiroki Nagakura, and Yudai Suwa for fruitful comments.
M.Z. is supported by JSPS KAKENHI Grant Numbers JP24H02245 and JP25K17383.
Numerical computations were carried out on the GPU cluster at the Center for Computational Astrophysics, National Astronomical Observatory of Japan.\\

\appendix
\section{Collisional spurious modes}\label{Sec:App}
In the case of $\bar{R}_0 = 10^{-4}$ where the far/gapped mode is dominant, we cannot reliably identify the damping near/gapless modes in the right panels of Fig.\,\ref{fig:DR_branch}.
This is because the roots are obscured by hundreds of spurious modes.

Recalling the phase space integrals in Eq.\,\eqref{eq:I_n_DR}, it has poles satisfying
\begin{equation}
    \omega - vk + \ii R_{\mathrm{EA}} = 0.
\end{equation}
The singular points just yield a branch cut on the Riemann surface analytically and do not prohibit the appearance of the roots.
However, in numerical integrations with discrete phase space grid, the integrals do not diverge, unless the grid point $(E_{\nu},v)$ accidentally satisfies the above singular condition.
Then, the discretized integrals introduce artificial poles, which can instead generate spurious roots in the dispersion relation \cite{Morinaga:2018}.

In the case of FFI and the traditional slow flavor instability, the existence of spurious modes stemming from the discrete angular integration has attracted attention \cite{Sarikas:2012,Morinaga:2018}.
However, it has been investigated in the absence of incoherent collisions, and the danger in the energy integration has been less explored.
In the previous studies, the branch cut is spanned over the real axis of wave frequency $\omega$ and $k$.
On the other hand, the collision terms bring decoherence directly, the branch cut extends along the imaginary orientation.

Focusing on the zero mode $k=0$, for instance, the phase space integral is singular at $\omega = - \ii R_{\mathrm{EA}}(E_{\nu})$.
In the continuous energy distributions, the branch cut is located from $\omega = -\ii 0$ for $E_{\nu} = 0$ to $\omega = -\ii\infty$ for $E_{\nu} = +\infty$.
In other words, the whole of negative imaginary axis in $\omega$ corresponds to the branch cut on the Riemann surface, and it is unfortunately reduced to the sequence of spurious modes.

\begin{figure}[t]
    \centering
    \includegraphics[width=1.\linewidth]{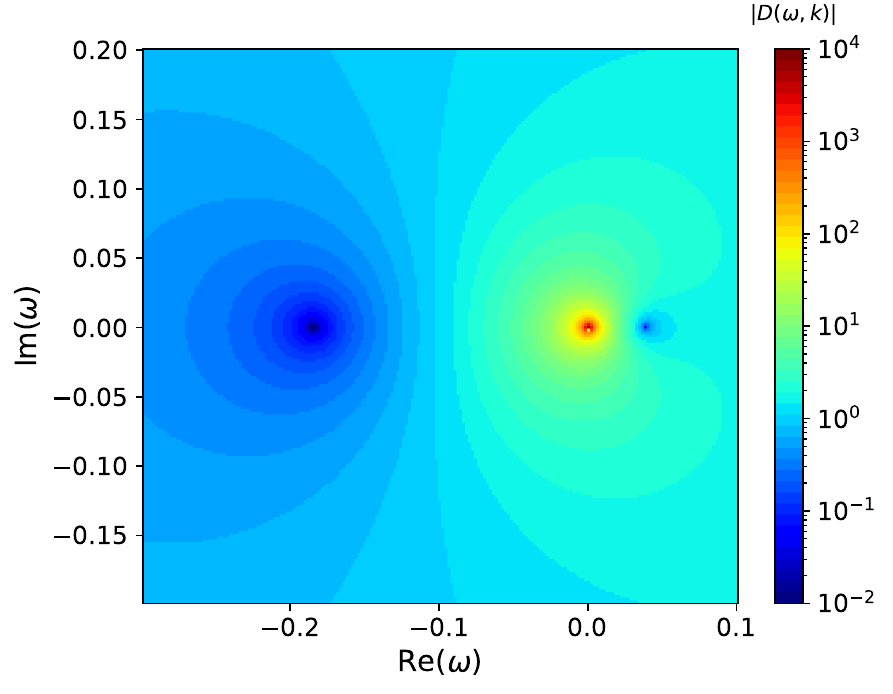}
    \includegraphics[width=1.\linewidth]{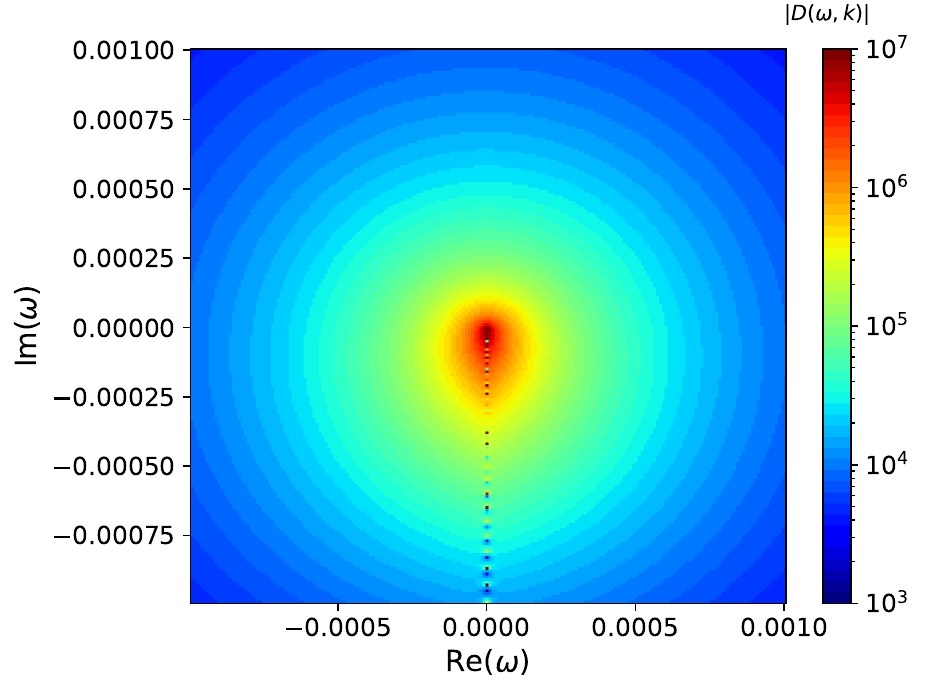}
    \caption{Absolute value of $\mathcal{D}(\omega,k)$ in the complex plane of $\omega$ for $k=0$.
    Top: the wide view where two roots with $\mathrm{Re}\,\omega\neq 0$ can be at least confirmed.
    Bottom: the focus view where the near/gapless modes would emerge around $\mathrm{Re}\,\omega\sim 0$.}
    \label{fig:Riemann}
\end{figure}
Go back to the actual situation in this work.
As mentioned in the above discussions, the near/gapless modes appear in the vicinity of the zero mode $k=0$.
In the case of $\bar{R}_0 = 10^{-5}$, they are growing modes and located on the positive side of the imaginary axis.
In contrast, in the case of $\bar{R}_0 = 10^{-4}$ where this numerical difficulty arises, they are expected to be damping modes and coexist with collisional spurious modes.
Figure\,\ref{fig:Riemann} displays the absolute value of $\mathcal{D}(\omega,k)$,
\begin{equation}
    \mathcal{D}(\omega,k) \equiv (I_0+1)(I_2-1) - I_1^2,
\end{equation}
for the zero mode $k=0$ on the complex plane of $\omega$.
Sufficiently small or exact zero value for $\mathcal{D}(\omega,k)$ expresses the roots in flavor instability.
The top panel shows the wide view, and the existence of two clear roots can be observed around $\mathrm{Re}\omega \sim -0.2$ and $0.06$.
They correspond to the growing far/gapped modes, as shown in Fig.\,\ref{fig:DR_branch}.
Magnifying the region around the origin, the sequence of poles mimicking a root can be seen along the negative direction of the imaginary axis.
The number of spurious modes is $2N_E = 200$ because neutrino and anti-neutrino sectors are accumulated.
The expected damping near/gapless modes could not be reliably isolated, and unfortunately we were unable to unravel them.
To obtain them adequately, we need to perform the phase space integration analytically even for the energy distributions.


\bibliography{3D_CFI}

@article{Abbar:2024a,
  title = {Physics-Informed Neural Networks for Predicting the Asymptotic Outcome of Fast Neutrino Flavor Conversions},
  author = {Abbar, Sajad and Wu, Meng-Ru and Xiong, Zewei},
  year = 2024,
  month = feb,
  journal = {Physical Review D},
  volume = {109},
  number = {4},
  pages = {043024},
  publisher = {American Physical Society},
  doi = {10.1103/PhysRevD.109.043024},
  urldate = {2026-11-24}
}

@article{Akaho:2024a,
  title = {Collisional and Fast Neutrino Flavor Instabilities in Two-Dimensional Core-Collapse Supernova Simulation with {{Boltzmann}} Neutrino Transport},
  author = {Akaho, Ryuichiro and Liu, Jiabao and Nagakura, Hiroki and Zaizen, Masamichi and Yamada, Shoichi},
  year = 2024,
  month = jan,
  journal = {Physical Review D},
  volume = {109},
  number = {2},
  pages = {023012},
  publisher = {American Physical Society},
  doi = {10.1103/PhysRevD.109.023012},
  urldate = {2024-11-18}
}

@article{Akaho:2026,
  title = {Bifurcated {{Impact}} of {{Neutrino Fast Flavor Conversion}} on {{Core-Collapse Supernovae Informed}} by {{Multiangle Neutrino Radiation Hydrodynamics}}},
  author = {Akaho, Ryuichiro and Nagakura, Hiroki and Iwakami, Wakana and Furusawa, Shun and Harada, Akira and Okawa, Hirotada and Matsufuru, Hideo and Sumiyoshi, Kohsuke and Yamada, Shoichi},
  year = 2026,
  month = may,
  journal = {Physical Review Letters},
  volume = {136},
  number = {19},
  pages = {191002},
  publisher = {American Physical Society},
  doi = {10.1103/fksy-1jtw},
  urldate = {2026-05-12}
}

@article{Bhattacharyya:2021,
  title = {Fast {{Flavor Depolarization}} of {{Supernova Neutrinos}}},
  author = {Bhattacharyya, Soumya and Dasgupta, Basudeb},
  year = 2021,
  month = feb,
  journal = {Physical Review Letters},
  volume = {126},
  number = {6},
  pages = {061302},
  publisher = {American Physical Society},
  doi = {10.1103/PhysRevLett.126.061302},
  urldate = {2021-03-19}
}

@article{Bhattacharyya:2022,
  title = {Elaborating the Ultimate Fate of Fast Collective Neutrino Flavor Oscillations},
  author = {Bhattacharyya, Soumya and Dasgupta, Basudeb},
  year = 2022,
  month = nov,
  journal = {Physical Review D},
  volume = {106},
  number = {10},
  pages = {103039},
  publisher = {American Physical Society},
  doi = {10.1103/PhysRevD.106.103039},
  urldate = {2022-12-29}
}

@article{Duan:2010,
  title = {Collective {{Neutrino Oscillations}}},
  author = {Duan, Huaiyu and Fuller, George M. and Qian, Yong-Zhong},
  year = 2010,
  month = oct,
  journal = {Annual Review of Nuclear and Particle Science},
  volume = {60},
  number = {1},
  pages = {569--594},
  publisher = {Annual Reviews},
  issn = {0163-8998},
  doi = {10.1146/annurev.nucl.012809.104524},
  urldate = {2020-06-18}
}

@article{Ehring:2023,
  title = {Fast Neutrino Flavor Conversion in Core-Collapse Supernovae: {{A}} Parametric Study in {{1D}} Models},
  shorttitle = {Fast Neutrino Flavor Conversion in Core-Collapse Supernovae},
  author = {Ehring, Jakob and Abbar, Sajad and Janka, Hans-Thomas and Raffelt, Georg and Tamborra, Irene},
  year = 2023,
  month = may,
  journal = {Physical Review D},
  volume = {107},
  number = {10},
  pages = {103034},
  publisher = {American Physical Society},
  doi = {10.1103/PhysRevD.107.103034},
  urldate = {2023-05-23}
}

@article{Ehring:2023a,
  title = {Fast {{Neutrino Flavor Conversions Can Help}} and {{Hinder Neutrino-Driven Explosions}}},
  author = {Ehring, Jakob and Abbar, Sajad and Janka, Hans-Thomas and Raffelt, Georg and Tamborra, Irene},
  year = 2023,
  month = aug,
  journal = {Physical Review Letters},
  volume = {131},
  number = {6},
  pages = {061401},
  publisher = {American Physical Society},
  doi = {10.1103/PhysRevLett.131.061401},
  urldate = {2023-08-23}
}

@article{Ehring:2026,
  title = {Gravitational-{{Wave Signatures}} of {{Nonstandard Neutrino Properties}} in {{Collapsing Stellar Cores}}},
  author = {Ehring, Jakob and Abbar, Sajad and Janka, Hans-Thomas and Raffelt, Georg and Nakamura, Ko and Kotake, Kei},
  year = 2026,
  month = jan,
  journal = {Physical Review Letters},
  volume = {136},
  number = {2},
  pages = {021201},
  publisher = {American Physical Society},
  doi = {10.1103/rv17-jm6g},
  urldate = {2026-04-10}
}

@article{Fernandez:2022,
  title = {Fast Flavor Instability in Hypermassive Neutron Star Disk Outflows},
  author = {Fern{\'a}ndez, Rodrigo and Richers, Sherwood and Mulyk, Nicole and Fahlman, Steven},
  year = 2022,
  month = nov,
  journal = {Physical Review D},
  volume = {106},
  number = {10},
  pages = {103003},
  publisher = {American Physical Society},
  doi = {10.1103/PhysRevD.106.103003},
  urldate = {2026-04-09}
}

@article{Fiorillo:2026c,
  title = {Dispersion Relation of the Neutrino Plasma: Unifying Fast, Slow, and Collisional Instabilities},
  shorttitle = {Dispersion Relation of the Neutrino Plasma},
  author = {Fiorillo, Damiano F. G. and Raffelt, Georg G.},
  year = 2026,
  month = jan,
  journal = {Journal of High Energy Physics},
  volume = {2026},
  number = {1},
  pages = {147},
  issn = {1029-8479},
  doi = {10.1007/JHEP01(2026)147},
  urldate = {2026-04-02},
  langid = {english}
}

@misc{Fiorillo:2026h,
  title = {Collective Flavor Conversion in Dense Neutrino Plasmas},
  author = {Fiorillo, Damiano F. G.},
  year = 2026,
  month = sep,
  number = {arXiv:2609.19256},
  eprint = {2609.19256},
  primaryclass = {hep-ph},
  publisher = {arXiv},
  doi = {10.48550/arXiv.2609.19256},
  urldate = {2026-09-18},
  archiveprefix = {arXiv}
}

@article{Fischer:2024,
  title = {Neutrinos and Nucleosynthesis of Elements},
  author = {Fischer, Tobias and Guo, Gang and Langanke, Karlheinz and {Mart{\'i}nez-Pinedo}, Gabriel and Qian, Yong-Zhong and Wu, Meng-Ru},
  year = 2024,
  month = may,
  journal = {Progress in Particle and Nuclear Physics},
  volume = {137},
  pages = {104107},
  issn = {0146-6410},
  doi = {10.1016/j.ppnp.2024.104107},
  urldate = {2025-12-26}
}

@article{Froustey:2025,
  title = {Predicting the Outcome of Collisional Neutrino Flavor Conversion},
  author = {Froustey, Julien},
  year = 2025,
  month = jul,
  journal = {Physical Review D},
  volume = {112},
  number = {2},
  pages = {023029},
  publisher = {American Physical Society},
  doi = {10.1103/syxg-sqfn},
  urldate = {2026-03-02}
}

@article{Froustey:2026,
  title = {Neutrino Flavor Instabilities in Neutron Star Mergers with Moment Transport: {{Slow}}, Fast, and Collisional Modes},
  shorttitle = {Neutrino Flavor Instabilities in Neutron Star Mergers with Moment Transport},
  author = {Froustey, Julien and Foucart, Francois and Hall, Christian and Kneller, James P. and Kundu, Debraj and Lin, Zidu and McLaughlin, Gail C. and Richers, Sherwood},
  year = 2026,
  month = mar,
  journal = {Physical Review D},
  volume = {113},
  number = {6},
  pages = {063050},
  publisher = {American Physical Society},
  doi = {10.1103/htq2-d1t7},
  urldate = {2026-08-18}
}

@article{Gogilashvili:2026,
  title = {Flavor Conversion Enhances or Suppresses Supernova Explodability Independent of the Progenitor Mass},
  author = {Gogilashvili, Mariam and Tamborra, Irene},
  year = 2026,
  month = aug,
  journal = {Physical Review D},
  volume = {114},
  number = {4},
  pages = {043054},
  publisher = {American Physical Society},
  doi = {10.1103/j5b2-5q2p},
  urldate = {2026-11-16}
}

@article{Gogilashvili:2026a,
  title = {Neutrino Flavor Conversion Shapes the Rate of Failed Core-Collapse Supernovae},
  author = {Gogilashvili, Mariam and Tamborra, Irene},
  year = 2026,
  month = sep,
  journal = {Physical Review D},
  volume = {114},
  number = {6},
  pages = {L061304},
  publisher = {American Physical Society},
  doi = {10.1103/pz3y-3lv5},
  urldate = {2026-11-16}
}

@article{Goimil-Garcia:2025a,
  title = {Steady State of Fast-Oscillating Neutrinos in an Inhomogeneous Medium},
  author = {{Goimil-Garc{\'i}a}, Manuel and Tamborra, Irene},
  year = 2025,
  month = nov,
  journal = {Physical Review D},
  volume = {112},
  number = {10},
  pages = {103011},
  publisher = {American Physical Society},
  doi = {10.1103/gdg9-rzns},
  urldate = {2026-11-16}
}

@article{Izaguirre:2017,
  title = {Fast {{Pairwise Conversion}} of {{Supernova Neutrinos}}: {{A Dispersion Relation Approach}}},
  shorttitle = {Fast {{Pairwise Conversion}} of {{Supernova Neutrinos}}},
  author = {Izaguirre, Ignacio and Raffelt, Georg and Tamborra, Irene},
  year = 2017,
  month = jan,
  journal = {Physical Review Letters},
  volume = {118},
  number = {2},
  pages = {021101},
  publisher = {American Physical Society},
  doi = {10.1103/PhysRevLett.118.021101},
  urldate = {2020-06-18}
}

@article{Johns:2023a,
  title = {Collisional {{Flavor Instabilities}} of {{Supernova Neutrinos}}},
  author = {Johns, Lucas},
  year = 2023,
  month = may,
  journal = {Physical Review Letters},
  volume = {130},
  number = {19},
  pages = {191001},
  publisher = {American Physical Society},
  doi = {10.1103/PhysRevLett.130.191001},
  urldate = {2023-05-09}
}

@misc{Johns:2023c,
  title = {Collisional Flavor Pendula and Neutrino Quantum Thermodynamics},
  author = {Johns, Lucas and Rodriguez, Santiago},
  year = 2023,
  month = dec,
  number = {arXiv:2312.10340},
  eprint = {2312.10340},
  primaryclass = {astro-ph, physics:hep-ph},
  publisher = {arXiv},
  doi = {10.48550/arXiv.2312.10340},
  urldate = {2024-07-25},
  archiveprefix = {arXiv}
}

@article{Johns:2025,
  title = {Neutrino {{Oscillations}} in {{Core-Collapse Supernovae}} and {{Neutron Star Mergers}}},
  author = {Johns, Lucas and Richers, Sherwood and Wu, Meng-Ru},
  year = 2025,
  journal = {Annual Review of Nuclear and Particle Science},
  volume = {75},
  pages = {399--423},
  publisher = {Annual Reviews},
  issn = {1545-4134},
  doi = {10.1146/annurev-nucl-121423-100853},
  urldate = {2026-09-17}
}

@article{Just:2022,
  title = {Fast Neutrino Conversion in Hydrodynamic Simulations of Neutrino-Cooled Accretion Disks},
  author = {Just, Oliver and Abbar, Sajad and Wu, Meng-Ru and Tamborra, Irene and Janka, Hans-Thomas and Capozzi, Francesco},
  year = 2022,
  month = apr,
  journal = {Physical Review D},
  volume = {105},
  number = {8},
  pages = {083024},
  publisher = {American Physical Society},
  doi = {10.1103/PhysRevD.105.083024},
  urldate = {2026-04-09}
}

@article{Kato:2024,
  title = {Collisional Flavor Swap with Neutrino Self-Interactions},
  author = {Kato, Chinami and Nagakura, Hiroki and Johns, Lucas},
  year = 2024,
  month = may,
  journal = {Physical Review D},
  volume = {109},
  number = {10},
  pages = {103009},
  publisher = {American Physical Society},
  doi = {10.1103/PhysRevD.109.103009},
  urldate = {2024-05-24}
}

@article{Li:2021,
  title = {Neutrino {{Fast Flavor Conversions}} in {{Neutron-Star Postmerger Accretion Disks}}},
  author = {Li, Xinyu and Siegel, Daniel M.},
  year = 2021,
  month = jun,
  journal = {Physical Review Letters},
  volume = {126},
  number = {25},
  pages = {251101},
  publisher = {American Physical Society},
  doi = {10.1103/PhysRevLett.126.251101},
  urldate = {2022-07-29}
}

@article{Lin:2023,
  title = {Collision-Induced Flavor Instability in Dense Neutrino Gases with Energy-Dependent Scattering},
  author = {Lin, Yu-Chia and Duan, Huaiyu},
  year = 2023,
  month = apr,
  journal = {Physical Review D},
  volume = {107},
  number = {8},
  pages = {083034},
  publisher = {American Physical Society},
  doi = {10.1103/PhysRevD.107.083034},
  urldate = {2023-09-04}
}

@article{Liu:2023,
  title = {Systematic Study of the Resonancelike Structure in the Collisional Flavor Instability of Neutrinos},
  author = {Liu, Jiabao and Zaizen, Masamichi and Yamada, Shoichi},
  year = 2023,
  month = jun,
  journal = {Physical Review D},
  volume = {107},
  number = {12},
  pages = {123011},
  publisher = {American Physical Society},
  doi = {10.1103/PhysRevD.107.123011},
  urldate = {2023-06-12}
}

@article{Liu:2023a,
  title = {Universality of the Neutrino Collisional Flavor Instability in Core-Collapse Supernovae},
  author = {Liu, Jiabao and Akaho, Ryuichiro and Ito, Akira and Nagakura, Hiroki and Zaizen, Masamichi and Yamada, Shoichi},
  year = 2023,
  month = dec,
  journal = {Physical Review D},
  volume = {108},
  number = {12},
  pages = {123024},
  publisher = {American Physical Society},
  doi = {10.1103/PhysRevD.108.123024},
  urldate = {2024-07-26}
}

@article{Liu:2024,
  title = {Muon-Induced Collisional Flavor Instability in Core-Collapse Supernova},
  author = {Liu, Jiabao and Nagakura, Hiroki and Akaho, Ryuichiro and Ito, Akira and Zaizen, Masamichi and Furusawa, Shun and Yamada, Shoichi},
  year = 2024,
  month = aug,
  journal = {Physical Review D},
  volume = {110},
  number = {4},
  pages = {043039},
  publisher = {American Physical Society},
  doi = {10.1103/PhysRevD.110.043039},
  urldate = {2024-08-29}
}

@article{Liu:2026,
  title = {Approximate Energy-Integration Method for Identifying Collisional Neutrino Flavor Instabilities},
  author = {Liu, Jiabao and Nagakura, Hiroki},
  year = 2026,
  month = jun,
  journal = {Physical Review D},
  volume = {113},
  number = {12},
  pages = {123047},
  publisher = {American Physical Society},
  doi = {10.1103/559p-9d8p},
  urldate = {2026-10-15}
}

@article{Liu:2026a,
  title = {Dynamical Equilibria of Fast Neutrino Flavor Conversion},
  author = {Liu, Jiabao and Johns, Lucas and Nagakura, Hiroki and Zaizen, Masamichi and Yamada, Shoichi},
  year = 2026,
  month = aug,
  journal = {Physical Review D},
  volume = {114},
  number = {4},
  pages = {L041306},
  publisher = {American Physical Society},
  doi = {10.1103/ydrn-np36},
  urldate = {2026-09-17}
}

@article{Lund:2025,
  title = {Angle-Dependent in {{Situ Fast Flavor Transformations}} in {{Post}}-Neutron-Star-Merger {{Disks}}},
  author = {Lund, Kelsey A. and Mukhopadhyay, Payel and Miller, Jonah M. and McLaughlin, G. C.},
  year = 2025,
  month = may,
  journal = {The Astrophysical Journal Letters},
  volume = {985},
  number = {1},
  pages = {L9},
  publisher = {The American Astronomical Society},
  issn = {2041-8205},
  doi = {10.3847/2041-8213/add0a7},
  urldate = {2025-09-12},
  langid = {english}
}

@article{Mori:2025,
  title = {Three-Dimensional Core-Collapse Supernova Models with Phenomenological Treatment of Neutrino Flavor Conversions},
  author = {Mori, Kanji and Takiwaki, Tomoya and Kotake, Kei and Horiuchi, Shunsaku},
  year = 2025,
  month = apr,
  journal = {Publications of the Astronomical Society of Japan},
  volume = {77},
  number = {2},
  pages = {L9-L15},
  issn = {2053-051X},
  doi = {10.1093/pasj/psaf007},
  urldate = {2025-09-12}
}

@article{Morinaga:2018,
  title = {Linear Stability Analysis of Collective Neutrino Oscillations without Spurious Modes},
  author = {Morinaga, Taiki and Yamada, Shoichi},
  year = 2018,
  month = jan,
  journal = {Physical Review D},
  volume = {97},
  number = {2},
  pages = {023024},
  publisher = {American Physical Society},
  doi = {10.1103/PhysRevD.97.023024},
  urldate = {2020-06-18}
}

@article{Nagakura:2024,
  title = {Neutron Star Kick Driven by Asymmetric Fast-Neutrino Flavor Conversion},
  author = {Nagakura, Hiroki and Sumiyoshi, Kohsuke},
  year = 2024,
  month = may,
  journal = {Physical Review D},
  volume = {109},
  number = {10},
  pages = {103017},
  doi = {10.1103/PhysRevD.109.103017},
  urldate = {2026-04-09}
}

@article{Nagakura:2025a,
  title = {Neutrino Flavor Instabilities in a Binary Neutron Star Merger Remnant: {{Roles}} of a Long-Lived Hypermassive Neutron Star},
  shorttitle = {Neutrino Flavor Instabilities in a Binary Neutron Star Merger Remnant},
  author = {Nagakura, Hiroki and Sumiyoshi, Kohsuke and Fujibayashi, Sho and Sekiguchi, Yuichiro and Shibata, Masaru},
  year = 2025,
  month = aug,
  journal = {Physical Review D},
  volume = {112},
  number = {4},
  pages = {043029},
  publisher = {American Physical Society},
  doi = {10.1103/kf8q-d67t},
  urldate = {2026-03-06}
}

@misc{Ng:2026,
  title = {Three-Flavor Neutrino {{Boltzmann}} Transport and Flavor Instabilities in Long-Lived Neutron-Star Merger Remnants},
  author = {Ng, Harry Ho-Yin and Liu, Jiabao and Offermans, Arthur and Yamada, Shoichi and Most, Elias R. and Rezzolla, Luciano},
  year = 2026,
  month = oct,
  doi = {10.48550/arXiv.2610.36486},
  urldate = {2026-10-28},
  langid = {english}
}

@article{Padilla-Gay:2022a,
  title = {Neutrino Fast Flavor Pendulum. {{II}}. {{Collisional}} Damping},
  author = {{Padilla-Gay}, Ian and Tamborra, Irene and Raffelt, Georg G.},
  year = 2022,
  month = nov,
  journal = {Physical Review D},
  volume = {106},
  number = {10},
  pages = {103031},
  publisher = {American Physical Society},
  doi = {10.1103/PhysRevD.106.103031},
  urldate = {2023-02-19}
}

@article{Pantaleone:1992,
  title = {Neutrino Oscillations at High Densities},
  author = {Pantaleone, James},
  year = 1992,
  month = aug,
  journal = {Physics Letters B},
  volume = {287},
  number = {1},
  pages = {128--132},
  issn = {0370-2693},
  doi = {10.1016/0370-2693(92)91887-F},
  urldate = {2020-06-18}
}

@article{Qiu:2025,
  title = {Neutrino {{Flavor Transformation}} in {{Neutron Star Mergers}}},
  author = {Qiu, Yi and Radice, David and Richers, Sherwood and Bhattacharyya, Maitraya},
  year = 2025,
  month = aug,
  journal = {Physical Review Letters},
  volume = {135},
  number = {9},
  pages = {091401},
  publisher = {American Physical Society},
  doi = {10.1103/h2q7-kn3v},
  urldate = {2025-09-12}
}

@article{Qiu:2025a,
  title = {Impact of Neutrino Flavor Conversions on Neutron Star Merger Dynamics, Ejecta, Nucleosynthesis, and Multimessenger Signals},
  author = {Qiu, Yi and Radice, David and Richers, Sherwood and Guercilena, Federico Maria and Perego, Albino and Bhattacharyya, Maitraya},
  year = 2025,
  month = dec,
  journal = {Physical Review D},
  volume = {112},
  number = {12},
  pages = {123039},
  doi = {10.1103/qckq-78gt},
  urldate = {2026-04-09}
}

@article{Raffelt:2013,
  title = {Axial {{Symmetry Breaking}} in {{Self-Induced Flavor Conversion of Supernova Neutrino Fluxes}}},
  author = {Raffelt, Georg and Sarikas, Srdjan and Seixas, David de Sousa},
  year = 2013,
  month = aug,
  journal = {Physical Review Letters},
  volume = {111},
  number = {9},
  pages = {091101},
  publisher = {American Physical Society},
  doi = {10.1103/PhysRevLett.111.091101},
  urldate = {2020-06-18}
}

@incollection{Raffelt:2026,
  title = {Neutrinos from Core-Collapse Supernovae},
  booktitle = {Encyclopedia of {{Particle Physics}} ({{First Edition}})},
  author = {Raffelt, Georg G. and Janka, Hans-Thomas and Fiorillo, Damiano F. G.},
  editor = {Fischer, Christian S.},
  year = 2026,
  month = jan,
  pages = {307--349},
  publisher = {Elsevier},
  address = {Oxford},
  doi = {10.1016/B978-0-443-26598-3.00065-1},
  urldate = {2026-11-16},
  isbn = {978-0-443-26599-0}
}

@article{Richers:2021,
  title = {Particle-in-Cell Simulation of the Neutrino Fast Flavor Instability},
  author = {Richers, Sherwood and Willcox, Don E. and Ford, Nicole M. and Myers, Andrew},
  year = 2021,
  month = apr,
  journal = {Physical Review D},
  volume = {103},
  number = {8},
  pages = {083013},
  publisher = {American Physical Society},
  doi = {10.1103/PhysRevD.103.083013},
  urldate = {2021-09-12}
}

@article{Richers:2022,
  title = {Code Comparison for Fast Flavor Instability Simulations},
  author = {Richers, Sherwood and Duan, Huaiyu and Wu, Meng-Ru and Bhattacharyya, Soumya and Zaizen, Masamichi and George, Manu and Lin, Chun-Yu and Xiong, Zewei},
  year = 2022,
  month = aug,
  journal = {Physical Review D},
  volume = {106},
  number = {4},
  pages = {043011},
  publisher = {American Physical Society},
  doi = {10.1103/PhysRevD.106.043011},
  urldate = {2022-09-29}
}

@incollection{Richers:2022b,
  title = {Fast {{Flavor Transformations}}},
  booktitle = {Handbook of {{Nuclear Physics}}},
  author = {Richers, Sherwood and Sen, Manibrata},
  editor = {Tanihata, Isao and Toki, Hiroshi and Kajino, Toshitaka},
  year = 2022,
  pages = {1--17},
  publisher = {Springer Nature},
  address = {Singapore},
  doi = {10.1007/978-981-15-8818-1_125-1},
  urldate = {2023-10-20},
  isbn = {978-981-15-8818-1},
  langid = {english}
}

@article{Richers:2024,
  title = {Asymptotic-State Prediction for Fast Flavor Transformation in Neutron Star Mergers},
  author = {Richers, Sherwood and Froustey, Julien and Ghosh, Somdutta and Foucart, Francois and Gomez, Javier},
  year = 2024,
  month = nov,
  journal = {Physical Review D},
  volume = {110},
  number = {10},
  pages = {103019},
  publisher = {American Physical Society},
  doi = {10.1103/PhysRevD.110.103019},
  urldate = {2025-02-14}
}

@article{Sarikas:2012,
  title = {Spurious Instabilities in Multiangle Simulations of Collective Flavor Conversion},
  author = {Sarikas, Srdjan and {de Sousa Seixas}, David and Raffelt, Georg},
  year = 2012,
  month = dec,
  journal = {Physical Review D},
  volume = {86},
  number = {12},
  pages = {125020},
  publisher = {American Physical Society},
  doi = {10.1103/PhysRevD.86.125020},
  urldate = {2020-06-18}
}

@article{Sawyer:2005,
  title = {Speed-up of Neutrino Transformations in a Supernova Environment},
  author = {Sawyer, R. F.},
  year = 2005,
  month = aug,
  journal = {Physical Review D},
  volume = {72},
  number = {4},
  pages = {045003},
  publisher = {American Physical Society},
  doi = {10.1103/PhysRevD.72.045003},
  urldate = {2020-06-18}
}

@article{Sawyer:2016,
  title = {Neutrino {{Cloud Instabilities Just}} above the {{Neutrino Sphere}} of a {{Supernova}}},
  author = {Sawyer, R. F.},
  year = 2016,
  month = feb,
  journal = {Physical Review Letters},
  volume = {116},
  number = {8},
  pages = {081101},
  publisher = {American Physical Society},
  doi = {10.1103/PhysRevLett.116.081101},
  urldate = {2020-06-18}
}

@article{Sigl:1993,
  ids = {Sigl:1993a},
  title = {General Kinetic Description of Relativistic Mixed Neutrinos},
  author = {Sigl, G. and Raffelt, G.},
  year = 1993,
  month = sep,
  journal = {Nuclear Physics B},
  volume = {406},
  number = {1},
  pages = {423--451},
  issn = {0550-3213},
  doi = {10.1016/0550-3213(93)90175-O},
  urldate = {2020-06-18}
}

@misc{Takahashi:2026,
  title = {Dynamic {{Competition}} of {{Fast}} and {{Collisional Neutrino Flavor Instabilities}} with {{Collisional Damping}} in {{Spatially Inhomogeneous Systems}}},
  author = {Takahashi, Shota and Nagakura, Hiroki and Zaizen, Masamichi and Kato, Chinami and Liu, Jiabao},
  year = 2026,
  month = may,
  number = {arXiv:2605.10435},
  eprint = {2605.10435},
  publisher = {arXiv},
  doi = {10.48550/arXiv.2605.10435},
  urldate = {2026-05-12},
  archiveprefix = {arXiv},
  langid = {english}
}

@article{Tamborra:2021,
  title = {New {{Developments}} in  {{Flavor Evolution}} of a {{Dense Neutrino Gas}}},
  author = {Tamborra, Irene and Shalgar, Shashank},
  year = 2021,
  month = sep,
  journal = {Annual Review of Nuclear and Particle Science},
  publisher = {Annual Reviews},
  issn = {0163-8998},
  doi = {10.1146/annurev-nucl-102920-050505},
  urldate = {2021-06-25}
}

@article{Tamborra:2025,
  title = {Neutrinos from Explosive Transients at the Dawn of Multi-Messenger Astronomy},
  author = {Tamborra, Irene},
  year = 2025,
  month = jun,
  journal = {Nature Reviews Physics},
  volume = {7},
  number = {6},
  pages = {285--298},
  publisher = {Nature Publishing Group},
  issn = {2522-5820},
  doi = {10.1038/s42254-025-00828-2},
  urldate = {2026-11-17},
  copyright = {2025 Springer Nature Limited},
  langid = {english}
}

@article{Urquilla:2026a,
  title = {Neutrino Transport and Flavor Instabilities in a Postmerger Disk},
  author = {Urquilla, Erick and Shankar, Swapnil and Kundu, Debraj and Froustey, Julien and Richers, Sherwood and Miller, Jonah M. and McLaughlin, Gail C. and Kneller, James P. and Foucart, Francois},
  year = 2026,
  month = oct,
  journal = {Physical Review D},
  volume = {114},
  number = {8},
  pages = {083059},
  publisher = {American Physical Society},
  doi = {10.1103/4nhz-gs2f},
  urldate = {2026-11-17}
}

@article{Volpe:2024,
  title = {Neutrinos from Dense Environments: {{Flavor}} Mechanisms, Theoretical Approaches, Observations, and New Directions},
  shorttitle = {Neutrinos from Dense Environments},
  author = {Volpe, M. Cristina},
  year = 2024,
  month = jun,
  journal = {Reviews of Modern Physics},
  volume = {96},
  number = {2},
  pages = {025004},
  publisher = {American Physical Society},
  doi = {10.1103/RevModPhys.96.025004},
  urldate = {2024-07-26}
}

@misc{Wang:2025,
  title = {The {{Effect}} of the {{Fast-Flavor Instability}} on {{Core-Collapse Supernova Models}}},
  author = {Wang, Tianshu and Burrows, Adam},
  year = 2025,
  month = mar,
  number = {arXiv:2503.04896},
  eprint = {2503.04896},
  primaryclass = {astro-ph},
  publisher = {arXiv},
  doi = {10.48550/arXiv.2503.04896},
  urldate = {2025-03-10},
  archiveprefix = {arXiv}
}

@misc{Wang:2025a,
  title = {The {{Effect}} of the {{Collisional Flavor Instability}} on {{Core-Collapse Supernova Models}}},
  author = {Wang, Tianshu and Nagakura, Hiroki and Johns, Lucas and Burrows, Adam},
  year = 2025,
  month = jul,
  number = {arXiv:2507.01100},
  eprint = {2507.01100},
  primaryclass = {astro-ph},
  publisher = {arXiv},
  doi = {10.48550/arXiv.2507.01100},
  urldate = {2025-07-03},
  archiveprefix = {arXiv}
}

@misc{Wang:2025b,
  title = {The {{Effect}} of the {{Fast-Flavor Instability}} on {{Core-Collapse Supernova Models}}: {{II}}. {{Quasi-Equipartition}} and the {{Impact}} of {{Various Angular Reconstruction Methods}}},
  shorttitle = {The {{Effect}} of the {{Fast-Flavor Instability}} on {{Core-Collapse Supernova Models}}},
  author = {Wang, Tianshu and Burrows, Adam},
  year = 2025,
  month = nov,
  number = {arXiv:2511.20767},
  eprint = {2511.20767},
  primaryclass = {astro-ph},
  publisher = {arXiv},
  doi = {10.48550/arXiv.2511.20767},
  urldate = {2025-11-27},
  archiveprefix = {arXiv}
}

@article{Wu:2021,
  title = {Collective Fast Neutrino Flavor Conversions in a {{1D}} Box: {{Initial}} Conditions and Long-Term Evolution},
  shorttitle = {Collective Fast Neutrino Flavor Conversions in a {{1D}} Box},
  author = {Wu, Meng-Ru and George, Manu and Lin, Chun-Yu and Xiong, Zewei},
  year = 2021,
  month = nov,
  journal = {Physical Review D},
  volume = {104},
  number = {10},
  pages = {103003},
  publisher = {American Physical Society},
  doi = {10.1103/PhysRevD.104.103003},
  urldate = {2021-11-17}
}

@article{Xiong:2023a,
  title = {Evaluating Approximate Asymptotic Distributions for Fast Neutrino Flavor Conversions in a Periodic {{1D}} Box},
  author = {Xiong, Zewei and Wu, Meng-Ru and Abbar, Sajad and Bhattacharyya, Soumya and George, Manu and Lin, Chun-Yu},
  year = 2023,
  month = sep,
  journal = {Physical Review D},
  volume = {108},
  number = {6},
  pages = {063003},
  publisher = {American Physical Society},
  doi = {10.1103/PhysRevD.108.063003},
  urldate = {2023-10-04}
}

@article{Xiong:2023b,
  title = {Collisional Flavor Instability in Dense Neutrino Gases},
  author = {Xiong, Zewei and Johns, Lucas and Wu, Meng-Ru and Duan, Huaiyu},
  year = 2023,
  month = oct,
  journal = {Physical Review D},
  volume = {108},
  number = {8},
  pages = {083002},
  publisher = {American Physical Society},
  doi = {10.1103/PhysRevD.108.083002},
  urldate = {2023-10-04}
}

@misc{Xiong:2026,
  title = {Neutrino Quantum Kinetics for Fast Flavor Conversion in a Time-Dependent Environment},
  author = {Xiong, Zewei and Wu, Meng-Ru},
  year = 2026,
  month = aug,
  number = {arXiv:2608.07773},
  eprint = {2608.07773},
  primaryclass = {astro-ph.HE},
  publisher = {arXiv},
  doi = {10.48550/arXiv.2608.07773},
  urldate = {2026-08-12},
  archiveprefix = {arXiv}
}

@article{Zaizen:2021a,
  title = {Nonlinear Evolution of Fast Neutrino Flavor Conversion in the Preshock Region of Core-Collapse Supernovae},
  author = {Zaizen, Masamichi and Morinaga, Taiki},
  year = 2021,
  month = oct,
  journal = {Physical Review D},
  volume = {104},
  number = {8},
  pages = {083035},
  publisher = {American Physical Society},
  doi = {10.1103/PhysRevD.104.083035},
  urldate = {2021-10-28}
}

@article{Zaizen:2023,
  title = {Simple Method for Determining Asymptotic States of Fast Neutrino-Flavor Conversion},
  author = {Zaizen, Masamichi and Nagakura, Hiroki},
  year = 2023,
  month = may,
  journal = {Physical Review D},
  volume = {107},
  number = {10},
  pages = {103022},
  publisher = {American Physical Society},
  doi = {10.1103/PhysRevD.107.103022},
  urldate = {2023-05-15}
}

@article{Zaizen:2025,
  title = {Spectral Diversity in Collisional Neutrino-Flavor Conversion: {{Flavor}} Equipartition or Swap},
  shorttitle = {Spectral Diversity in Collisional Neutrino-Flavor Conversion},
  author = {Zaizen, Masamichi},
  year = 2025,
  month = may,
  journal = {Physical Review D},
  volume = {111},
  number = {10},
  pages = {103029},
  publisher = {American Physical Society},
  doi = {10.1103/PhysRevD.111.103029},
  urldate = {2025-05-16}
}

@misc{Zaizen:2026,
  title = {Fast {{Neutrino-Flavor Conversion}} with {{Attenuation}} and {{Global Lepton Gradient}}},
  author = {Zaizen, Masamichi and Nagakura, Hiroki},
  year = 2026,
  month = apr,
  number = {arXiv:2604.13617},
  eprint = {2604.13617},
  primaryclass = {astro-ph},
  publisher = {arXiv},
  doi = {10.48550/arXiv.2604.13617},
  urldate = {2026-04-17},
  archiveprefix = {arXiv}
}

@misc{Zaizen:2026d,
  title = {Movie for {{Figs}}.3 and 7 of {{Collisional}} Dynamics of Neutrino Flavor Conversion: Angular Structures, Inhomogeneity, and Decoherence},
  author = {Zaizen, Masamichi},
  year = 2026,
  url = {https://mzizn.github.io/movie/3D-CFI/3D-CFI.html}
}

\end{document}